\documentclass[letterpaper,11pt,reqno]{amsart}

\makeatletter
\usepackage{amssymb}
\usepackage{latexsym}
\usepackage{amsbsy}
\usepackage{amsfonts}
\usepackage{caption}
\usepackage{subcaption}
\usepackage{hyperref}
\usepackage{graphicx}
\usepackage{enumerate}
\usepackage{enumitem}
\usepackage{color}
\usepackage{algorithm}% http://ctan.org/pkg/algorithm
\usepackage[noend]{algpseudocode}% http://ctan.org/pkg/algorithmicx
\usepackage[round]{natbib}   % omit 'round' option if you prefer square brackets

\usepackage{tikz}

\def\marginpar#1{\ignorespaces}

\newcounter{procedure}
\makeatletter
\newenvironment{procedure}[1][htb]{%
    \let\c@algorithm\c@procedure
    \renewcommand{\ALG@name}{Procedure}% Update algorithm name
    \begin{algorithm}[#1]%
    }{\end{algorithm}
}
\makeatother

\DeclareMathOperator\cov{Cov}
\DeclareMathOperator\var{Var}
\DeclareMathOperator\corr{Corr}
\DeclareMathOperator\cord{CORD}
\DeclareMathOperator\argmin{argmin}
\DeclareMathOperator\argmax{argmax}

\newtheorem{theorem}{Theorem}

\newtheorem{assumption}{Assumption}
\newtheorem{criterion}{Criterion}
\newtheorem{rot}{Rule}
\newtheorem{lemma}{Lemma}
\makeatother
\begin{document}
\title[Correlation Blockmodel Clustering]{Asset Selection via Correlation Blockmodel Clustering}

\author[Wenpin Tang]{{Wenpin} Tang}
\thanks{Tang gratefully acknowledges financial support through an NSF grant DMS-2113779 and through a start-up grant at Columbia University.
Zhou gratefully acknowledges financial supports through a start-up grant at Columbia University and through the Nie Center for Intelligent Asset Management.}
\address{Department of Industrial Engineering and Operations Research, Columbia University.}
\email{wt2319@columbia.edu}

\author[Xiao Xu]{{Xiao} Xu}
\address{Department of Industrial Engineering and Operations Research, Columbia University.}
\email{xx2167@columbia.edu}

\author[Xun Yu Zhou]{{Xun Yu} Zhou}
\address{Department of Industrial Engineering and Operations Research, Columbia University.}
\email{xz2574@columbia.edu}

\date{\today}
\begin{abstract}
We aim to cluster financial assets in order to identify a small set of stocks to approximate the level of diversification of the whole universe of stocks.
We develop a data-driven approach to clustering based on a correlation blockmodel
in which assets in the same cluster are highly correlated with each other and, at the same time, have the same correlations with all other assets.
We devise an algorithm to detect the clusters, with theoretical analysis and practical guidance.
Finally, we conduct an empirical analysis to verify the performance of the algorithm.
\end{abstract}

\maketitle

\setcounter{tocdepth}{1}
%\tableofcontents
%-------------------------------------------------------------------------------------------------
\section{Introduction}

\quad The modern portfolio theory was pioneered by \cite{Mar52, Mar59}, in which the key insights are diversification and risk-return tradeoff.
One drawback of applying Markowitz's mean-variance portfolio selection approach na\"ively is to include {\it all} the available assets for allocation.
So in the case of S\&P 500, for example, an investor would need to invest in all these 500 stocks in her portfolio.
This is simply impossible for small investors or small fund managers.\footnote{Some investment experts suggest that 30 stocks be the maximum number of stocks in a retail investor's stock portfolio (``How Many Stocks Should Be in a Portfolio?'', Zacks, 2019, Accessed January 5th, 2021 \url{https://finance.zacks.com/many-stocks-should-portfolio-4782.html}).}
Seasoned investors such as Warren Buffet do not hold many stocks either.\footnote{ Between Berkshire Hathaway and New England Asset Management, Buffet holds 49 stocks in total, with about 92\% of the portfolio concentrating in 15 stocks, and 78\% in just five stocks (based on holdings as of September 30, 2020, reported in Berkshire Hathaway and New England Asset Management’s 13F filings on November 16, 2020).}
Even gigantic funds such as Vanguard and BlackRock do not include, even though they could, all the stocks in their portfolios.
Practically, managing too many stocks is costly and prone to mismanagement.
According to a Morningstar article, ``when you lose your focus and move outside your circle of competence, you lose your competitive advantage.''\footnote{``How Many Stocks Diversify Unsystematic Risk?'', Morningstar, Accessed January 5th, 2021.
\url{http://news.morningstar.com/classroom2/course.asp?docId=145385&page=4}.} Technically, including too many stocks increases both the odds of overfitting and the difficulty in computing efficient allocation strategies (e.g., \cite{DGU09}).
One way to address this issue is to add a regularization term or a cardinality constraint in the Markowitz mean-variance optimization model  (\cite{Faaland1974, DeMiguel2009, Gao2013, BD09, HSX15}).
This approach imposes sparsity on the number of assets in the portfolio; however, the regularization itself does not take diversification into account, and thus the set of stocks selected may contain concentration risk.

\quad Since the main reason to include all the stocks is to diversify, we have the following natural question: How can we select a {\it much} smaller subset of the whole universe of stocks that achieves a {\it sufficient}  level of diversification?\footnote{If this question can be satisfactorily answered, we can then apply Markowitz's mean-variance model to this small set of stocks to get an efficient portfolio.
In other words, we can decompose the Markowitz model into two stages: asset {\it selection} and asset {\it allocation}.}
\cite{Reilly2012} states that ``about 90\% of the maximum benefit of diversification was derived from portfolios of 12 to 18 stocks''.
\cite{Mar52} suggests a simple rule of thumb for selecting stocks that one should try to ``diversify across industries because firms in different industries, especially industries with different economic characteristics, have lower covariances than firms within an industry''.
In practice, stock selection is often based on factors such as sector rotation and macroeconomic indicators and is subjective to each investor.
However, this approach relies on the taxonomy of sectors and macroeconomic analysis published by certain organizations, which again may be subject to different interpretations and may contain biases.
If, as noted above, the primary goal of stock selection is to achieve sufficient diversification to which asset correlations are the key, approaches focusing directly on asset correlations are more appropriate and more innately fitting for the subsequent asset allocation.
A promising such approach is {\it clustering} based on correlation networks.
Specifically, one first groups or clusters all the assets in a correlation network and then selects one or a few ``representative" assets in each group,
resulting in a subset of a much smaller number of assets.

\quad Since the seminal work of \cite{Man99}, correlation networks have been widely used as a tool to study the correlation structure of financial assets.
In a correlation network, financial assets are modeled as nodes, which are then connected by edges representing the correlations between their returns.
Clustering analysis is conducted on correlation networks, and clusters are often compared with traditional industry classifications \citep{Rosen2006, Musmeci2015a}.
One line of the clustering research is simply to understand the market structure without involving portfolio selection; see \cite{MS99, Das03} with $k$-means algorithm, \cite{GA00} with hierarchical clustering, and \cite{Man99, TA05} with network filtering.
Another line of research is to utilize the revealed market structure to construct portfolios.
For instance, \cite{Ren2005} creates clusters based on a simple threshold rule and constructs an optimal portfolio of sub-portfolios, each of which is an equally-weighted portfolio of all stocks in the same cluster.
Based on just the structure of the correlation networks, \cite{Pozzi2013a} build portfolios consisting of stocks in the center and on the periphery of the networks.
For various studies  applying the same idea of ``clustering analysis and portfolio construction using representative sub-portfolios'', see e.g., \cite{Nanda2010}, \cite{Zhan2015}, \cite{Marvin2015}, \cite{DM16}, \cite{R17}, \cite{Leon2017}, \cite{Korzeniewski2018}.
Most recently, \cite{Puerto2020} propose a unified framework that pursues high within-cluster correlations while at the same time optimizing the allocations of the representative stocks.

\quad Despite their popularity in the machine learning community, the $k$-means algorithm \citep{Lloyd1982} and the similar $k$-medoids algorithm (or Partitioning Around Medoids, ``PAM'', \citealp{Kaufman1990}) have found only limited applications to finance.
\cite{Marvin2015} applies $k$-medoids to cluster financial assets, albeit based on financial ratios of companies instead of return time series.
\cite{Musmeci2015a} compare  $k$-medoids  with other clustering methods along with the industry classification.
\cite{He2007} and \cite{Nakagawa2019a} apply $k$-means and $k$-medoids to financial time series data for a different purpose: they group series of returns in order to predict future returns.

\quad Most of the aforementioned clustering methods applied to portfolio selection are heuristics, hence often difficult to interpret.
In this paper, we propose a new, interpretable, data-driven approach to correlation
network clustering and provide a systematic solution for selecting well-diversified stocks.
Our clustering is based on the following two criteria:

\begin{criterion}
\label{criterion:nb1}
Financial assets in the same group have high correlations.
\end{criterion}

\begin{criterion}
\label{criterion:nb2}
Financial assets in the same group have similar correlations with {\it all} other assets.
\end{criterion}

\quad Criterion \ref{criterion:nb1} is self-evident.
Assets with high correlations may perform well simultaneously at one time and plummet simultaneously at another time;
so we cluster them into the same group.
The next important question is how to select the ``representative" assets in each group of a clustering.
Practitioners often choose the best assets in each group according to their own performance metrics.
However, this does not guarantee that the assets selected in each group are ``optimal'' especially in terms of their relationship with other assets outside of the group.
Motivated by the mean-variance portfolio theory, we propose to cluster in such a way that any two assets in the same group have similar correlations to {\it all} others in the stock universe, which underlines  Criterion \ref{criterion:nb2}.
So any two assets in the same group are interchangeable in terms of their correlations with other assets. Consequently, one needs to only choose some idiosyncratic characteristics, such as volatility or Sharpe ratio, in selecting which asset to be included in the portfolio.
This makes the choice of representative assets from each cluster simple and transparent.
The idea, though very natural, seems missing in the literature.

\quad The purpose of this paper is to develop a new financial clustering approach, taking {\it both} criteria into account.
We propose a {\em correlation blockmodel} to capture Criterion \ref{criterion:nb2}.
This formulation is inspired by the problem of community detection in stochastic blockmodel \citep{Abbe17} and block covariance model \citep{Bunea2016, BGL20}.
In the model, the return of any asset in the same group is expressed as the sum of a common latent factor and an uncorrelated random noise.
As such, any two assets in the same group have the same correlations to all others.
Criterion \ref{criterion:nb1} is then used to calibrate a threshold hyperparameter that controls how variables are grouped together.
We devise an algorithm -- called ACC (Asset Clustering through Correlation) -- to recover the clusters of the blockmodel in polynomial time.

\quad The contributions of this work are as follows.
First, to our best knowledge,  this paper is the first to implement both criteria (especially Criterion 2) in financial asset clustering to capture the notion of diversification and the first to utilize the correlation blockmodel to formalize the implementation.
This provides interpretability of our clustering approach from the portfolio theory point of view.
Second, we lay a rigorous foundation for the clustering algorithm from both algorithmic and statistical perspectives.
In particular, we provide a statistical guarantee for the algorithm which can account for the possible heavy-tailed data intrinsic to financial time series.
This estimate requires a delicate analysis owing to the heavy tails and is new and interesting in its own right.
Moreover, we propose a hyperparameter tuning procedure in the clustering algorithm, taking both criteria into account.
The information limit of the blockmodel narrows down the search for the hyperparameter, while Criterion \ref{criterion:nb1} is used to cross-validate.
Finally, we conduct an extensive empirical study on the S\&P 500 stocks by selecting 15 to 25 stocks at a time via clustering and constructing portfolios using the selected stocks.
For comparison, we select stocks from clusters created by the popular $k$-medoids clustering algorithm and clusters based on S\&P's sector and industry classification.
We also consider the set of all S\&P 500 sector ETFs, each of which represents a different sector in the S\&P 500 Index.
For all these groups of stocks, we employ and compare three asset allocation strategies: risk parity, minimum-variance, and Markowitz's mean-variance optimal allocation.
The results show that the portfolios constructed using our ACC algorithm outperform the benchmark -- the S\&P 500 ETF -- significantly.
The portfolios based on ACC clusters also perform favorably compared to all other portfolios, especially when portfolios are readjusted infrequently.

% \quad In \cite{Bunea2016} (which is an unpublished, earlier version of \cite{BGL20}), an algorithm called CORD (CORrelation Difference) is presented to generate the clustering, and an R-package of the same name is created.
\quad In \cite{Bunea2016} (which is an unpublished, earlier version of \cite{BGL20}), an algorithm is presented to recover the clusters under the same correlation blockmodel.
Our results and algorithm differ significantly from \cite{Bunea2016} in the following aspects.
First, while \cite{Bunea2016} briefly demonstrate their model by applying the clustering algorithm to stock data in their numerical experiments, we are motivated by the modern portfolio theory to construct and justify the model and offer a theoretical interpretation of the model related to diversification based on the two criteria.
Second, we employ a different tuning procedure in order to incorporate Criterion \ref{criterion:nb1}.
Third, the underlying distribution is restricted to Gaussian in \cite{Bunea2016}, while we consider a more general range of distributions that encompass the heavy-tails prevalent in financial data.
Lastly, we conduct portfolio construction and extensive backtesting, which is not the primary focus of \cite{Bunea2016}.

\quad The remainder of the paper is organized as follows.
In Section \ref{sc2}, we present the correlation blockmodel and the ACC algorithm and state the main theoretical results.
Section \ref{sc4} provides an empirical analysis.
We conclude with a few remarks in Section \ref{sc5}.
All the proofs are contained in the Appendix.

%-------------------------------------------------------------------------------------------------
\section{Correlation blockmodel and clustering algorithm}
\label{sc2}
\bigskip
We first collect some notations that will be used throughout this paper.
All vectors are column vectors unless stated otherwise.
\begin{itemize}[label = {--}, itemsep = 3 pt]
\item
We use bold case letters, e.g., $\pmb{X}$, to denote matrices.
\item
For a vector $x$, $|x|$ is the Euclidean norm of $x$.
\item
For a vector $x$ (resp. a matrix $\pmb{X}$), $x^{\top}$ (resp. $\pmb{X}^{\top}$) is the transpose of $x$ (resp. $\pmb{X}$).
\item
For a set $A$, $|A|$ is the number of elements in $A$.
\item
For a random variable $X$, $\mathbb{E}(X)$ is the expectation of $X$, and $\var(X)$ the variance of $X$.
\item
For two random variables $X$ and $Y$, $\cov(X,Y):= \mathbb{E}[(X-\mathbb{E}(X))(Y-\mathbb{E}(Y))]$ is the covariance between $X$ and $Y$, and
$$\corr(X,Y):= \frac{\mathbb{E}[(X-\mathbb{E}(X))(Y-\mathbb{E}(Y))]}{\sqrt{\var(X)\var(Y)}},$$
is the Pearson correlation coefficient between $X$ and $Y$.
\item
$a$ and $b$ have the same order of magnitude, denoted by $a \asymp b$, if and only if $c\leq a/b\leq C$ for some $c, C>0$, as $a,b\rightarrow \infty$.
\end{itemize}

%-------------------------------------------------------------------------------------------------
\subsection{Model setup}
\label{sc21}
Assume that there are $d$ financial assets, indexed by $[d]:=\{1, \ldots, d\}$.
For $i \in [d]$, let $X_i$ be the return of asset $i$, and $X^{*}_i: = (X_i - \mathbb{E}(X_i))/\sqrt{\var(X_i)}$ be the standardized return.
As mentioned in the introduction, one of our goals is to cluster the returns $X = (X_1, \ldots, X_d)^\top$ in such a way that $X_i$ and $X_j$ belong to the same group if and only if they have the same correlations with {\it all} other returns, i.e., $\corr(X_i, X_l) = \corr(X_j, X_l)$ for $l \ne i, j$.
This amounts to finding a partition $G = \{G_1, \ldots, G_K\}$ of $[d]$, or a map of membership assignment $z: [d] \to [K]$ so that
\begin{equation*}
X_i \mbox{ belongs to group } k \Longleftrightarrow  i \in G_k  \Longleftrightarrow z(i) = k.
\end{equation*}
The sets $G_1, \ldots, G_K$ are called the {\it blocks}, or the groups of the partition $G$, which define an equivalence relation
$i \stackrel{G}{\sim} j$ if and only if  $i, j \in G_k$ for some $k \in [K]$, or simply $z(i) = z(j)$.
Similarly, $i \stackrel{G}{\nsim} j$ if and only if there is no block $G_k$ that contains both $i$ and $j$, or $z(i) \ne z(j)$.

\quad We now introduce the correlation blockmodel, in which $X_i$'s in the same group can be decomposed as the sum of a common latent factor and an uncorrelated random fluctuation.
Precisely, the standardized returns are represented as
\begin{equation}
\label{eq:CBM}
X^{*}_i = F_{z(i)} + U_i, \quad i \in [d],
\end{equation}
where
\begin{itemize}[itemsep = 3 pt]
\item
$F = (F_1, \ldots, F_K)$ are the latent factors with $\mathbb{E}(F_k) = 0$ for each $k\in[K]$;
\item
$U = (U_1, \ldots, U_d)$ are idiosyncratic fluctuations
with $\mathbb{E}(U_i) = 0$ for each $i\in[d]$, $\cov(U_i, U_j) = 0$ for $i \ne j$, and $\cov(F_{k}, U_i) = 0$ for each $k, i$.
\end{itemize}
It is easy to see that for $i \stackrel{G}{\sim} j$ and $l \ne i, j$, $\corr(X_i, X_l) = \mathbb{E}(F_{z(i)}F_{z(l)}) = \mathbb{E}(F_{z(j)}F_{z(l)}) = \corr(X_j, X_l)$.
Moreover, let $\sigma_k^2: = \var(F_k)$ be the variance of the latent factor underlying group $k \in [K]$.
By definition \eqref{eq:CBM}, $\var(U_i) = 1 - \sigma_k^2$ if $i \in G_k$.
This implies that the signal-to-noise ratios are the same for all standardized returns belonging to the same group.

\quad Given such a correlation blockmodel, the question is to infer the block structure -- the partition $G$ from the correlation matrix $\pmb{\rho}: = \mathbb{E}(X^{*} X^{*\top})$, where $X^{*} = (X^{*}_1, \ldots, X^{*}_d)^\top$ are standardized returns.
Denote $\pmb{\Sigma}_F: = \mathbb{E}(F F^{\top})$ and $\pmb{\Sigma}_{U}: = \mathbb{E}(U U^{\top})$ as the covariance matrices of $F$ and $U$ respectively.
Here $\pmb{\Sigma}_{U}$ is diagonal, since $\cov(U_i, U_j) = 0$ for $i \ne j$.
Let $\pmb{Z}:= (1_{\{z(i) = k\}})_{(i,k)\in [d] \times [K]}$ be the membership matrix.
The correlation matrix $\pmb{\rho}$ is then expressed as
\begin{equation}
\label{eq:corrrho}
\pmb{\rho} = \pmb{Z} \pmb{\Sigma}_F \pmb{Z}^{\top} + \pmb{\Sigma}_{U}.
\end{equation}
Note that the partition $G$ which satisfies \eqref{eq:CBM} or \eqref{eq:corrrho} may not be unique.
This could be due to overly granular partitions splitting the set of returns with the same latent factor further into smaller groups, or due to different latent factors that share the same covariances with all other latent factors.
A natural way to address this problem is to look for the {\it coarsest} partition, which has the least number of clusters.
Since the partition order is a partial order, the coarsest partition is, in general, still not necessarily unique.
Nevertheless, the following result ensures a unique coarsest partition $G^{\star}$ for the correlation blockmodel \eqref{eq:CBM}.
\begin{theorem}
\label{thm:uniquepart}
Let $\pmb{\rho}$ be a correlation matrix.
Then there is a unique coarsest partition $G^{\star}$ such that $\pmb{\rho} = \pmb{Z} \pmb{\Pi} \pmb{Z}^{\top} + \pmb{\Gamma}$ for some membership matrix $\pmb{Z}$ associated with $G^{\star}$, some matrix $\pmb{\Pi}$, and some diagonal matrix $\pmb{\Gamma}$.
Moreover, the partition $G^{\star}$ is defined by the equivalence relation
\begin{equation}
\label{eq:eqvclass}
i \stackrel{G^{\star}}{\sim} j \quad \mbox{if and only if} \quad \max_{l \ne i,j} |\rho_{il} - \rho_{jl}| = 0.
\end{equation}
\end{theorem}

\quad The proof is deferred to Appendix A.1.
Theorem \ref{thm:uniquepart} shows that the coarsest partition $G^{\star}$ is well-defined; so the clusters of the correlation blockmodel \eqref{eq:corrrho} are identifiable.
We note that the coarsest partition could potentially group two clusters controlled by different factors as one, if the factors have the same covariances with all other factors.
This is not a problem for our purposes, as this partition would still satisfy Criterion \ref{criterion:nb2}.
In the remainder of this work, we aim to recover or to estimate the partition $G^{\star}$ from historical data.

%-------------------------------------------------------------------------------------------------
\subsection{The PARTITION procedure}
\label{sc22}
According to Theorem \ref{thm:uniquepart}, two financial asset returns $X_i$ and $X_j$ belong to different clusters in $G^{*}$ if and only of $\max_{l \ne i,j} |\rho_{il} - \rho_{jl}| > 0$.
This observation motivates the definition of a dissimilarity measure between assets $i$ and $j$ -- correlation difference (CORD, \cite{Bunea2016}):
\begin{equation}
\label{eq:CORD}
\cord(i,j): = \max_{l \ne i, j} |\rho_{il} - \rho_{jl}|, \quad i,j \in [d].
\end{equation}
This measure quantifies the dissimilarity between two assets in terms of their respective correlations with {\it all} other assets.
Consider a set of financial asset returns that includes $X_i$ and some other returns $Y_1, Y_2, \ldots$.
If we position $X_i$ at the top, the corresponding row in the covariance matrix of these returns is:\[\Sigma_i = (\var(X_i), \cov(X_i, Y_1), \cov(X_i, Y_2), \ldots).\] If we have $\cord(i,j)=0$, then $\cov(X_j, Y_k) = c\cov(X_i, Y_k)$ for all $k=1,2,\ldots$, where $c = \sqrt{\var(X_j)/\var(X_i)}$ is a constant.
So if we replace $X_i$ with $X_j$ in the set of asset returns, the first row (and column) of the covariance matrix will only be rescaled by a constant factor $c$, except for the variance term, which will be scaled by $c^2$.
Then in a minimum-variance portfolio with no short selling, which optimizes the weights to minimize the portfolio variance based on the covariance matrix, the assets $i$ and $j$ are interchangeable up to a constant factor.\footnote{Recall that a no-short-selling minimum variance portfolio of a set of assets with covariance matrix $\pmb{\Sigma}$ is one with asset weights $\pmb{w}$ that solves the optimization problem:\begin{align*}
        \min_{\pmb{w}} \quad &\pmb{w}^\top \pmb{\Sigma}\pmb{w}\\
        \text{subject to}\quad &\pmb{w}^\top \pmb{1} = 1\\
        &\pmb{w}\geq 0.
    \end{align*}}
This interchangeability also leads to the following result, which will be proved in Appendix A.2.

\begin{theorem}
    \label{thm:global_minimum_variance}
    Under the correlation blockmodel with a coarsest partition $G^* = \{G_1, G_2, \ldots, G_K\}$, construct a minimum variance portfolio by choosing one asset from each cluster:
    \[P_J := \{J(1), J(2), \ldots, J(K)\},\] where $J(k)\in{G_k}$, for $k=1,\ldots,K$.
    Among all such portfolios $P_J$, the portfolio with the lowest variance is the one consists of the asset with the lowest variance in each cluster: $\argmin_J{\var(P_J)} = \{J^*(1), J^*(2), \ldots, J^*(K)\}$ where
    \begin{equation}
        J^*(k) = \argmin_{j\in G_k}{\var({X_j})}, \quad \forall k=1,\ldots, K.
    \end{equation}
\end{theorem}

\quad This theorem gives guidance on how to choose an asset from each cluster in order to attain the minimum variance among all possible minimum variance portfolios.
Arguably, this selection approach aligns with Markowitz's original notion that diversification can be measured by variance minimization.

\quad Next, we discuss how to derive clustering from data.
Assume that the financial asset returns are observed over $n$ periods.
For $r \in [n]$, let $X^r = (X^r_1, \ldots, X^r_d)^\top$ be the asset returns in period $r$,
and $X^{*r} = (X^{*r}_1, \ldots, X^{*r}_d)^\top$ be the corresponding standardized returns.
Denote by $\pmb{X}^{*}$ the $n \times d$ matrix whose row $r$ is $X^{*r}$.
Also assume that $X^1, \ldots, X^n$ are independent and identically distributed (i.i.d.) so that
$X^{*1}, \ldots, X^{*n}$ are i.i.d. copies of $X^{*}$ defined by \eqref{eq:CBM}.
The goal is to estimate the cluster partition $G^{\star}$ from the sample correlation matrix $\widehat{\pmb \rho}$ given by
\begin{equation}
\label{eq:srho}
\widehat{\pmb \rho}: = \frac{1}{n-1} (\pmb{X}^{*})^{\top} \pmb{X}^{*} = \frac{1}{n-1} \sum_{r = 1}^n X^{*r}(X^{*r})^{\top}.
\end{equation}
Define the sample correlation difference $\widehat{\cord}$ by
\begin{equation}
\label{eq:sCORD}
\widehat{\cord}(i,j):= \max_{l \ne i, j} |\widehat{\rho}_{il} - \widehat{\rho}_{jl}|, \quad i,j \in [d].
\end{equation}

\quad Given the sample correlation differences, the clusters can be recovered through an iterative procedure that takes a dissimilarity matrix $\pmb{D}$ and a threshold parameter $\varepsilon$ as inputs.
This procedure, which we call PARTITION, is a generalization of the CORD algorithm (\cite{Bunea2016}) and is described as Procedure \ref{proc:partition}.
The main idea of the PARTITION procedure is that two assets $i$ and $j$ should belong to the same cluster if their dissimilarity, denoted by $\pmb{D}(i,j)$, is small, e.g., below a threshold $\varepsilon > 0$.
In each iteration, the procedure identifies a new cluster by finding the most similar pair of assets, i.e., with the smallest $\pmb{D}(i,j)$.
If $\pmb{D}(i,j)$ between these two assets is lower than the predetermined threshold $\varepsilon$,
then the two assets act as the core of the cluster, and all other assets that are similar to {\it either} of the core assets are included in the cluster.
Otherwise, any one of the two assets is singled out as its own cluster.
Let us note that the PARTITION procedure does not require as input the number of clusters $K$, which is determined via the threshold $\varepsilon$.
In our method, we will use the sample correlation difference $\widehat{\cord}$ between assets as the input to recover the clusters under the correlation blockmodel.
We will explain the method to determine the appropriate value for the threshold $\varepsilon$ in the upcoming sections.

\begin{procedure}
    \caption{Partition}
    \label{proc:partition}
    \begin{algorithmic}
    \Procedure{PARTITION}{$\pmb{D}$, $\varepsilon$}\Comment{$\pmb{D}$ is a given dissimilarity matrix; $\varepsilon>0$}
        \State Initialization: $S \longleftarrow [d]$, $l \longleftarrow 0$.
        \While{$S \ne \emptyset$}
            \State $l \longleftarrow l+1$
            \If{$|S| = 1$}
                \State $\widehat{G}_l \longleftarrow S$
            \EndIf
            \If{$|S| > 1$}
                \State $(i_l, j_l) \longleftarrow \argmin_{i,j \in S, i \ne j} \pmb{D}(i,j)$
                \If{$\pmb{D}(i_l, j_l) > \varepsilon$}
                \State $\widehat{G}_l \longleftarrow \{i_l\}$
                \Else  \State $\widehat{G}_l \longleftarrow \left\{k \in S: \min\left(\pmb{D}(i_l, k), \pmb{D}(j_l, k)\right) \le \varepsilon \right\}$
                \EndIf
            \EndIf
            \State $S \longleftarrow S \setminus \widehat{G}_l$
        \EndWhile
        \State \Return $\widehat{G} = \{\widehat{G}_1, \widehat{G}_2, \ldots \}$
    \EndProcedure
    \end{algorithmic}
  \end{procedure}

\quad We now study the statistical property of the PARTITION procedure when applied to the sample correlation difference.
To do this, we need the following assumption on the distribution of the asset returns.
\begin{assumption}
\label{assump:alphaexp}
The correlation matrix $\pmb{\rho}$ is non-singular, and the vector $\pmb{\rho}^{-1/2}X^{*}$ is $\alpha$-sub-exponential, $\alpha \in (0,2]$;
that is, there exists $L > 0$ such that
\begin{equation*}
||\pmb{\rho}^{-1/2}X^{*}||_{\psi_{\alpha}} \le L,
\end{equation*}
where $||Z||_{\psi_{\alpha}}:=\sup_{||\omega||_2 = 1} \inf\{s > 0: \mathbb{E}(e^{(|Z^{\top} \omega|/s)^{\alpha}}) \le 2 \}$
is the $\alpha$-Orlicz norm of $Z \in \mathbb{R}^d$.
\end{assumption}

\quad The non-singularity amounts to the non-existence of redundant securities, i.e., there does not exist any asset whose return is a linear combination of those of the other assets in the universe.
The $\alpha$-sub-exponential distribution was introduced in \cite{KR61} to characterize heavy-tailed random variables.
The special cases $\alpha = 2$ and $\alpha = 1$ correspond to the sub-Gaussian and sub-exponential variables respectively.
The lower $\alpha$ is, the more heavy-tailedness is allowed.
Assumption \ref{assump:alphaexp} is motivated by the stylized fact that financial assets often have heavy-tailed returns; see, e.g., \cite{Cont01}.

\quad The following result provides the statistical guarantee for the PARTITION procedure, along with guidance for the choice of the threshold $\varepsilon$.
\begin{theorem}
\label{thm:recovery}
Under Assumption \ref{assump:alphaexp}, there exist numerical constants $c_1, c_2 > 0$ independent of $n$ and $d$, such that if
$\min_{i \stackrel{G^{\star}}{\nsim} j} \cord(i,j) > \varepsilon$ and
\begin{equation}
\label{eq:varepsLB}
\varepsilon \ge 2L^2 \left(c_1 \sqrt{\frac{\log d}{n}} + c_2 \frac{(\log d)^{2/\alpha}}{n} \right),
\end{equation}
then the PARTITION procedure with inputs $\widehat{\cord}$ and $\varepsilon$ outputs $\widehat{G} = G^{\star}$ with probability $1 - 4/d$.
\end{theorem}

\quad The proof of Theorem \ref{thm:recovery} is deferred to Appendix A.3.
Theorem \ref{thm:recovery} implies that under a cluster separation condition and when the number of variables $d$ is large,
the PARTITION procedure recovers the clusters with high probability if
the threshold $\varepsilon$ is roughly of order $\max(\sqrt{\log d \, /n}, (\log d)^{2/\alpha}/n)$.
When $d=500$ for instance, this probability is 99.2\%.

\quad Notice that $\sqrt{\log d \, /n}$ dominates $(\log d)^{2/\alpha}/n$ if $n > (\log d)^{\frac{4}{\alpha}-1}$.
In practice, the number of observations $n$ is the same order as the number of assets $d$.
So as $n \asymp d \rightarrow \infty$, the right hand side of \eqref{eq:varepsLB} is dominated by $\sqrt{\log d \, /n}$.
However, the comparison of these two terms is sensitive to the value of $\alpha$ when $n \asymp d$ and both are finite.
For instance, consider a universe of $d = 500$ financial assets.
The table below displays the values of $n$ above which $\sqrt{\log d \, /n}>(\log d)^{2/\alpha}/n$ for different $\alpha$.
Therefore, it is important to accurately estimate $\alpha$ in order to determine which of the two terms, $\sqrt{\log d \, /n}$ and $(\log d)^{2/\alpha}/n$, dominates.
\begin{table}[h]
\begin{center}
\begin{tabular}{ |c|c|c|c|c|c|c|c|c|}
 \hline
 $\alpha$ & 0.25 & 0.5 & 0.75 & 1.0 & 1.25 & 1.5 & 1.75 & 2.0  \\ \hline
 $n$ & $7.97 \times 10^{11}$ & $3.58 \times 10^5$ & $2.74 \times 10^3$ & $240.02$ & $55.66$ & $21.01$ & $10.47$ & $6.21$  \\
 \hline
\end{tabular}
\smallskip
\caption{Values of $n$ above which $\sqrt{\log d \,/n}>(\log d)^{2/\alpha}/n$ for $d = 500$.}
\end{center}
\end{table}

%-------------------------------------------------------------------------------------------------
\subsection{Tuning the threshold $\varepsilon$}
\label{sc23}
The effectiveness of the PARTITION procedure depends on the threshold $\varepsilon$, which in turn determines the number of clusters $K$.
For instance, a low threshold, e.g., $\varepsilon = 0$, leads to many singleton clusters due to noise in the observations, while a high threshold, e.g., $\varepsilon = 2$, results in a single cluster because $\widehat{\cord}$, which is the maximum among differences between sample correlations, has a maximum value of $2$.
Theorem \ref{thm:recovery} provides a part of the guidance for choosing a suitable $\varepsilon$. Here, we propose a data-driven approach to tune the hyperparameter $\varepsilon$ based on the following three rules of thumb.

\quad First of all, according to Theorem \ref{thm:recovery}, the PARTITION procedure recovers the partition $G^{\star}$ if
$\varepsilon$ is of order $L^2\max(\sqrt{\log d \, /n}, (\log d)^{2/\alpha}/n)$.
This criterion gives a reasonable range for the choice of $\varepsilon$.
\begin{rot}
\label{rule:nb1}
The search range for the threshold $\varepsilon$ is determined as follows:
\begin{enumerate}[itemsep = 3 pt]
\item
If $n > (\log d)^{\frac{4}{\alpha}-1}$, then the range is set to be $[a,b] \times L^2\sqrt{\frac{\log d}{n}}$;
\item
If $n  \le(\log d)^{\frac{4}{\alpha}-1}$, then the range is set to be $[a,b] \times L^2\frac{(\log d)^{2/\alpha}}{n}$.
\end{enumerate}
Here $a, b$ are user-defined parameters.
\end{rot}

\quad With the sample correlation difference as input, the PARTITION procedure captures Criterion \ref{criterion:nb2}.
However, it does not take into account Criterion \ref{criterion:nb1} -- financial assets in the same cluster are highly correlated.
To incorporate this criterion, we propose to calibrate the value of $\varepsilon$ by considering the intra-cluster correlations.
To be more precise, let $\widehat{G}_{\varepsilon}$ be the output given by the PARTITION procedure with a threshold $\varepsilon$.
Define the intra-cluster correlation $\widehat{\rho}^{ave}_{\varepsilon}$ of $\widehat{G}_{\varepsilon}$ by
\begin{equation}
\label{eq:intracor}
\widehat{\rho}^{ave}_{\varepsilon}:= \frac{\sum_{i<j}1\Big(i \stackrel{\widehat{G}_{\varepsilon}}{\sim} j\Big) \widehat{\rho}_{ij}}{\sum_{i<j}1\Big(i \stackrel{\widehat{G}_{\varepsilon}}{\sim} j\Big)},
\end{equation}
where $\widehat{\rho}_{ij}$ is the sample correlation between asset returns $i$ and $j$ given by \eqref{eq:srho}.
The goal is to select the threshold $\varepsilon$ that gives the maximal intra-cluster correlation.
\begin{rot}
\label{rule:nb2}
Let $\mathcal{T}$ be the range specified in Rule \ref{rule:nb1}.
We choose
\begin{equation*}
\varepsilon_{\mathcal{T}}:= \argmax_{\varepsilon \in \mathcal{T}}\widehat{\rho}^{ave}_{\varepsilon} .
\end{equation*}
\end{rot}

\quad One disadvantage of applying Rule \ref{rule:nb2} na\"ively is that it is biased towards granular partitions with very few but high intra-cluster correlations.
Specifically, consider an extreme case where each asset forms its own cluster except for two assets $i$ and $j$ with correlation $\rho_{ij} = 1$, which form one cluster $\{i,j\}$.
By definition \eqref{eq:intracor}, the average intra-cluster correlation under this partition is $\widehat{\rho}^{ave} = \rho_{ij} = 1$.
However, such a partition that only groups two assets $i$ and $j$ together is not very informative, despite indeed being optimal under Rule \ref{rule:nb2}.
To regularize Rule \ref{rule:nb2}, we propose to put constraints on the number of clusters.
That is, we set a range for the number of clusters.
Thresholds resulting in too many or too few clusters are discarded, and then the one with the highest intra-cluster correlation is selected.
This results in the final rule:
\begin{rot}
\label{rule:nb3}
Let $\mathcal{T}$ be the range specified in Rule \ref{rule:nb1}, and $\mathcal{U}$ be a user-defined range for the number of clusters.
We choose
\begin{equation*}
\varepsilon_{\mathcal{T U}}:= \argmax_{\varepsilon \in \mathcal{T},\, |\widehat{G}_{\varepsilon}| \in \mathcal{U}} \widehat{\rho}^{ave}_{\varepsilon} .
\end{equation*}
\end{rot}

In solving the above maximization problem, the grid search is performed on the search range, split into $N$ grids, where $N$ is a user-defined parameter for the search.

\subsection{Estimation of heavy-tailedness}
\label{sec:heavy_tailedness_estimation}
By Rule \ref{rule:nb1}, the heavy-tailedness $\alpha$ together with some constant $L$ that depends on $\alpha$ determines the range in which we search for the threshold $\varepsilon$.
Thus, we need to estimate the parameter $\alpha$ and constant $L$, which encode the heavy-tailed nature of the returns $X^{*}$.

\quad First, \cite{Vladimirova2020} prove that Assumption \ref{assump:alphaexp} is equivalent to:
\begin{align*}
    \exists L>0 \text{ such that} \quad \mathbb{P}\left( \left|\left(\pmb{\rho}^{-1/2}X^{*}\right)^\top w\right|> t\right) \le 2 \exp\left(-(t/L)^{\alpha}\right) \quad \\
    \mbox{for } \forall t \ge 0 \mbox{ and } \forall w\in\mathbb{R}^d, ||w||_2=1,
\end{align*}
where the $L$ values are the same as in Assumption \ref{assump:alphaexp}.

\quad In order to facilitate the estimation of the tail parameters $L$ and $\alpha$, we further restrict $w$ to singleton vectors\footnote{The random variables in $\pmb{\rho}^{-1/2}X^{*}$ are uncorrelated. If we further assume that they are independent, then the $w$ leading to the heaviest tail should only select the one variable with the heaviest tail.}, i.e., vectors with only one element being $1$ while the rest being $0$.
This transforms the assumption to, for some $\alpha\in(0,2]$:
\begin{equation}
\label{eq:alphaexpbis}
\exists L>0 \text{ such that} \quad \mathbb{P}\left( \left|\left(\pmb{\rho}^{-1/2}X^{*}\right)_r\right|> t\right) \le 2 \exp\left(-(t/L)^{\alpha}\right) \quad \mbox{for } \forall t \ge 0 \mbox{ and } \forall r \in [d],
\end{equation}
where $(\pmb{\rho}^{-1/2}X^{*})_r$ is the $r^{th}$ coordinate of the random vector $\pmb{\rho}^{-1/2}X^{*}$.
Recall that the smaller $\alpha$ is, the more heavy-tailedness is allowed in the distribution.
So if $X^{*}$ satisfies \eqref{eq:alphaexpbis} for some $\alpha>0$, it also does for all $\alpha' \in (0, \alpha)$.
Similarly, for any fixed $\alpha$, if the inequality in \eqref{eq:alphaexpbis} holds for some $L$, it also holds for any $L'>L$.
Thus, we aim to find the largest $\alpha$ such that \eqref{eq:alphaexpbis} holds for all $r\in[d]$, and the smallest $L$ for such $\alpha$.
Notice that \eqref{eq:alphaexpbis} is trivially satisfied for small $t$, since when $t < L(\log2)^{\frac{1}{\alpha}}$, the right hand side is greater than $1$.
This means that Assumption \ref{assump:alphaexp} only controls the tail distribution of $|(\pmb{\rho}^{-1/2}X^{*})_r|$.
Given some $\alpha$ and $L$, consider the tail distribution characterized by the following survival function:
\begin{equation}
\label{eq:alphaexptaildist}
    \mathbb{P}(Y > t) = 2\exp\left(- (t/L)^{\alpha}\right) \quad \mbox{for } t \ge L(\log2)^{\frac{1}{\alpha}}.
\end{equation}
Loosely speaking, this distribution concerns the \textit{boundary} of the condition \eqref{eq:alphaexpbis}.
If for each coordinate $r$ in the random vector, we can find suitable $\alpha_r$ and $L_r$ such that the distribution \eqref{eq:alphaexptaildist} fits the tail observations of the vector $Y_r: = |(\pmb{\rho}^{-1/2}X^{*})_r|$, then $\alpha^{*}:=\min_{r \in [d]} \alpha_r$ and $L^{*}:=\max_{r \in [d]} L_r$ estimate the largest $\alpha$ and the smallest $L$ that satisfy the inequality \eqref{eq:alphaexpbis} for all $r\in[d]$.

\quad To estimate $\alpha$ and $L$ in \eqref{eq:alphaexptaildist}, we employ the idea in \cite{Gardes2008}.
The quantile function corresponding to \eqref{eq:alphaexptaildist} is $q(p):= \inf\{s > 0: \mathbb{P}(Y \le s) > p\} = L\left(\log\frac{2}{1-p}\right)^{\frac{1}{\alpha}}$.
Taking the log on both sides, we have
\begin{equation*}
    \log q(p) = \frac{1}{\alpha}\log\log\frac{2}{1-p}+\log L, \quad p \in (0,1),
\end{equation*}
which shows an affine relationship between $\log q(p)$ and $\log \log(2/(1-p))$ with the slope $1/\alpha$ and intersection $\log L$.
So we can apply linear regression to estimate the slope $1/\alpha$ and the corresponding constant $L$ using the tail observations.
Specifically, assume that we have $n$ i.i.d. observations ordered as $Y_r(1)\leq Y_r(2)\leq\ldots\leq Y_r(n)$.
Consider the largest $k$ observations $Y_r(n-j)$ for $1\leq j\leq k$.
Each of these observations approximates the quantile $(n-j)/n = 1-j/n$ (the largest observation $Y_r(n)$ corresponds to $q(1)$ and is not included).
Thus, we use the slope from the linear regression of $\log Y_r(n-j)$ against $\log\log(2n/j)$, for $1\leq j\leq k$, as an estimate of $1/\alpha$ and use the intersection as an estimate of $\log L$.

\subsection{The ACC algorithm}
Summarizing the above, we now describe the complete algorithm that recovers clusters from the raw input $\pmb{X}$, which we call the ACC (Asset Clustering through Correlation) algorithm and present as Algorithm \ref{algo:ACC}.

\begin{algorithm}[!htb]
    \caption{Asset Clustering through Correlation (ACC)}
    \label{algo:ACC}
    \begin{algorithmic}
      \Procedure{ACC}{$\pmb{X}$, $a$, $b$, $n_g$, $\mathcal{U}$, $k$} \Comment{Input: returns $\pmb{X}\in\mathbb{R}^{n\times d}$, search range $[a, b]$, number of grids $n_g$, range of clusters $\mathcal{U}$, number of large observations $k$}
      \State $\pmb{X}^*\longleftarrow \frac{\pmb{X} - \text{mean}(\pmb{X})}{\text{std}(\pmb{X})}$\Comment{Standardized returns; mean and std are column-wise}
      \State $\widehat{\pmb \rho} \longleftarrow \frac{1}{n-1} (\pmb{X}^{*})^{\top} \pmb{X}^{*}$
      \State $\widehat{\cord}(i,j)\longleftarrow \max_{l \ne i, j} |\widehat{\rho}_{il} - \widehat{\rho}_{jl}|, \quad \forall i,j \in [d]$
      \For{$i\in[d]$}
          \State $Y_i\longleftarrow|(\widehat{\pmb \rho}^{-1/2}\pmb{X}^{*})_r|\in\mathbb{R}^n$
          \State Sort $Y_i$ such that $Y_i[1]\leq Y_i[2]\leq \ldots \leq Y_i[n]$
          \State Slope $s$, intersection $a$ $\longleftarrow$ LinearRegression($\log Y_i[n-k : n-1] \sim \log\log(2n/[1:k])$)
          \State $\alpha_i\longleftarrow 1/s$, $L_i\longleftarrow \exp(a)$
      \EndFor
      \State $\alpha\longleftarrow\min_{i\in[d]}\alpha_i$
      \State $L\longleftarrow\max_{i\in[d]}L_i$
      \If{$n > (\log d)^{\frac{4}{\alpha}-1}$}
          \State Range $\mathcal{T}\longleftarrow[a,b] \times L^2\sqrt{\frac{\log d}{n}}$
      \Else
          \State Range $\mathcal{T}\longleftarrow[a,b] \times L^2\frac{(\log d)^{2/\alpha}}{n}$
      \EndIf
      \State Uniformly divide range $\mathcal{T}$ into $n_g$ grids
      \For{$\varepsilon$ in the $n_g$ grids of $\mathcal{T}$}
          \State $\widehat{G}_{\varepsilon} \longleftarrow \text{PARTITION}(\widehat{\cord}, \varepsilon)$
          \If{$|\widehat{G}_{\varepsilon}|\in\mathcal{U}$}
              $\widehat{\rho}^{ave}_{\varepsilon} \longleftarrow \frac{\sum_{i<j}1\Big(i \stackrel{\widehat{G}_{\varepsilon}}{\sim} j\Big) \widehat{\rho}_{ij}}{\sum_{i<j}1\Big(i \stackrel{\widehat{G}_{\varepsilon}}{\sim} j\Big)}$
          \Else
              \State $\widehat{\rho}^{ave}_{\varepsilon}\longleftarrow -\infty$.
          \EndIf
      \EndFor
      \State $\varepsilon_{\mathcal{T}}\longleftarrow\argmax_{\varepsilon \in \mathcal{T}}\widehat{\rho}^{ave}_{\varepsilon}$
      \State \Return $\text{PARTITION}(\widehat{\cord}, \varepsilon_{\mathcal{T}})$
    \EndProcedure
    \end{algorithmic}
  \end{algorithm}

\quad The algorithmic complexity of ACC is polynomial in both the number of assets $d$ and the number of observations $n$.
Specifically, we have the following theorem.
\begin{theorem}
\label{thm:complexity}
The ACC algorithm requires at most $\mathcal{O}(nd^2 + d^3)$ arithmetic operations.
\end{theorem}
We prove Theorem \ref{thm:complexity} in Appendix A.4.
%-------------------------------------------------------------------------------------------------
\section{Empirical analysis}
\label{sc4}

In this section, we report the results of empirical experiments applying the ACC algorithm to financial time series data.
Specifically, we cluster the stocks in the S\&P 500 universe using our ACC algorithm, together with two benchmarks: the $k$-medoids algorithm and the single-linkage hierarchical clustering algorithm, and analyze the quality of such clustering results.
Then based on the clustering results, we construct stock portfolios using three allocation strategies: the risk parity strategy, the minimum variance strategy, and Markowitz’s mean-variance strategy, and compare the performances of these portfolios.
We include additional benchmark portfolios based on the GICS sector and industry group classification and also portfolios of S\&P 500 sector ETFs.
These benchmarks will be explained in detail in Section \ref{sc44}.

%--------------------------------------------------------
\subsection{Data preparation}
\label{sc41}
We take the constituents of the S\&P 500 as the universe.
The data is obtained from Compustat through Wharton Research Data Services (WRDS), which consists of
\begin{itemize}[itemsep = 3 pt]
\item
the daily closing prices of the constituents;
\item
the historical constituents data; and
\item
the daily closing S\&P 500 total return index with dividends reinvested,
\end{itemize}
between January 1996 and January 2020.

\quad We conduct clustering and backtesting for the period between February 2001 and January 2020.
Clustering and portfolios are calculated on the first trading day of each month on the S\&P 500 constituent stocks.
Specifically, at the end of the first trading day of each month, we choose the stocks that are in the S\&P 500 index according to the historical constituents data.
Of all the current constituents, we discard stocks with less than five years of history and those with more than $5\%$ missing data in the past $n=500$ days.
If the same company has multiple classes of stocks in the S\&P 500 index (e.g., Alphabet Inc's GOOG and GOOGL),
we only choose the class with the longest history.
The numbers of eligible stocks remaining after the above filtering range between $465$ and $488$ over the backtesting period.
For these eligible stocks, any missing prices are linearly interpolated using the previous and subsequent prices.
Then, clusters are estimated based on the daily returns of the past $n =500$ trading days.
A smaller set of stocks is then selected, and portfolios are constructed using different allocation strategies.
We defer the details of portfolio construction to Section \ref{sc44}.

%--------------------------------------------------------

%--------------------------------------------------------
\subsection{Clustering procedure}
\label{sc43}
We now give a detailed description of how we use the ACC algorithm.
On the first trading day of each month, the ACC algorithm is applied to the sample correlation matrix in the backward 500-trading-day window for valid constituent stocks as described in Section \ref{sc41}.
The following highlights more specifics:
\begin{itemize}[itemsep = 3 pt]
\item
The heavy-tailedness parameter $\alpha$ and constant $L$ are estimated by the approach in Section \ref{sec:heavy_tailedness_estimation}, where we choose $k:=n/4 = 125$.
\item
The search range for the threshold parameter $\varepsilon$ is set by Rule \ref{rule:nb1} with $a = 0.1$, $b = 10$ and $N = 100$.
That is, if $n > \log(d)^{\frac{4}{\alpha}-1}$, then the range is $\mathcal{T} = [0.1, 10] \times  L^2\sqrt{\frac{\log(d)}{n}} $; otherwise $\mathcal{T} = [0.1, 10] \times L^2\frac{\log(d)^\frac{2}{\alpha}}{n}$.
We cap the upper bound of the search range to 2.
The grid search is then performed on the search range with the range split into $100$ grids.
\item
The number of clusters in Rule \ref{rule:nb3} is restricted between $15$ and $25$, i.e. $\mathcal{U} = [15, 25]$.
\end{itemize}

\begin{figure}[!htb]
    \centering
    \includegraphics[width=0.72\textwidth]{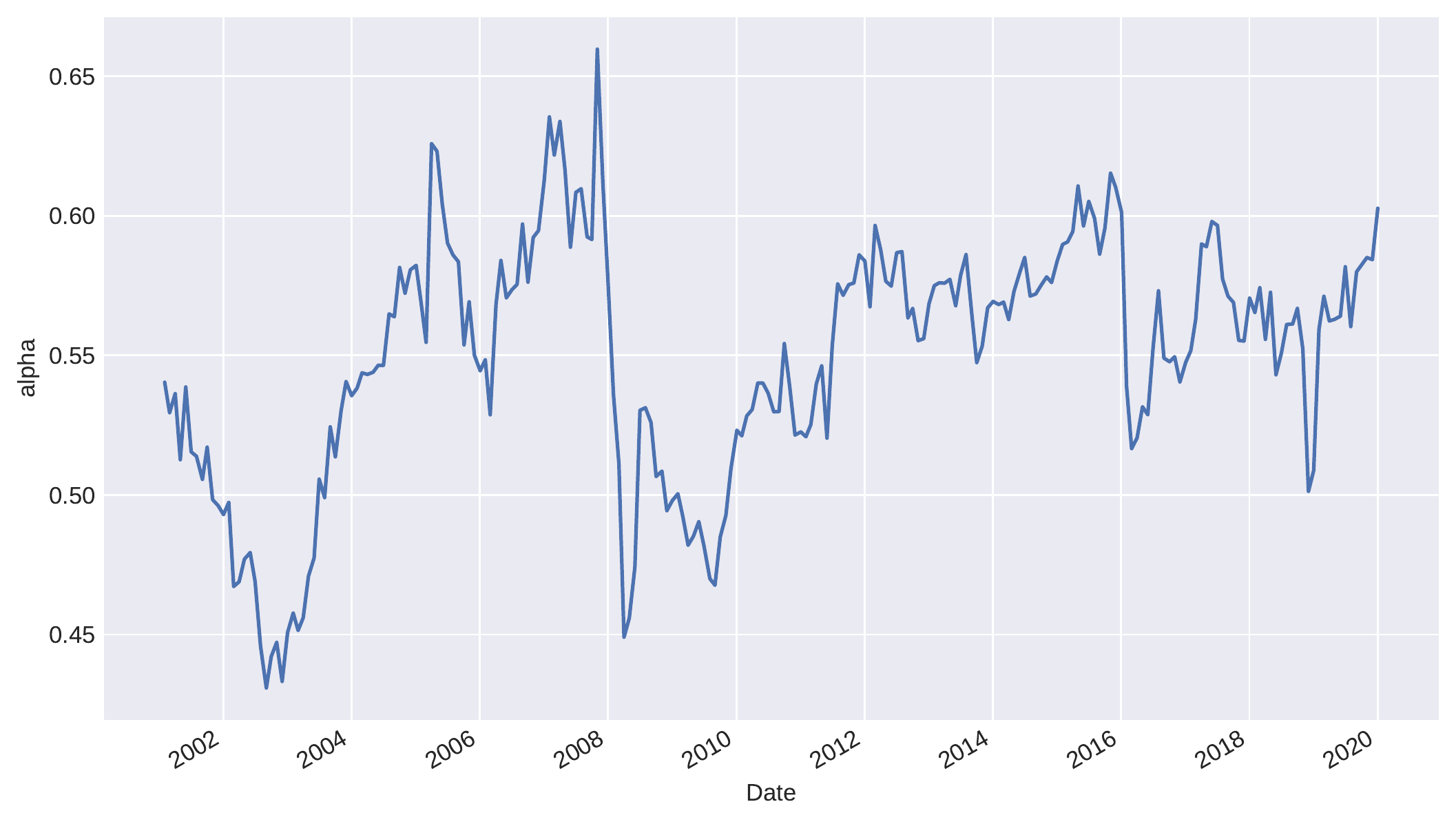}
    \caption{Estimated $\alpha$ values by month via 500-trading-day windows.}
    \label{fig:alpha_by_month}
\end{figure}

\quad From February 2001 to January 2020, a total of 228 partitions are constructed, one in each month.
The estimated $\alpha$ values estimated at the beginning of each month are plotted in Figure \ref{fig:alpha_by_month}.
The S\&P 500 data exhibits notable heavy tails, with $\alpha$ values lower than $1$.
The estimated value of the constant $L$ ranges between $0.6$ and $0.8$ over time.
It is also worth pointing out that the value of $\alpha$ drops significantly between 2008 and 2010, arguably due to more extreme returns observed during the 2008 financial crisis.\footnote{This is consistent with the increased tail risk measured by VaR observed in stock daily returns during the financial crisis; see, e.g., \cite{Chaudhury2014}.}

\begin{figure}[!htb]
    \centering
    \includegraphics[width=0.72 \textwidth]{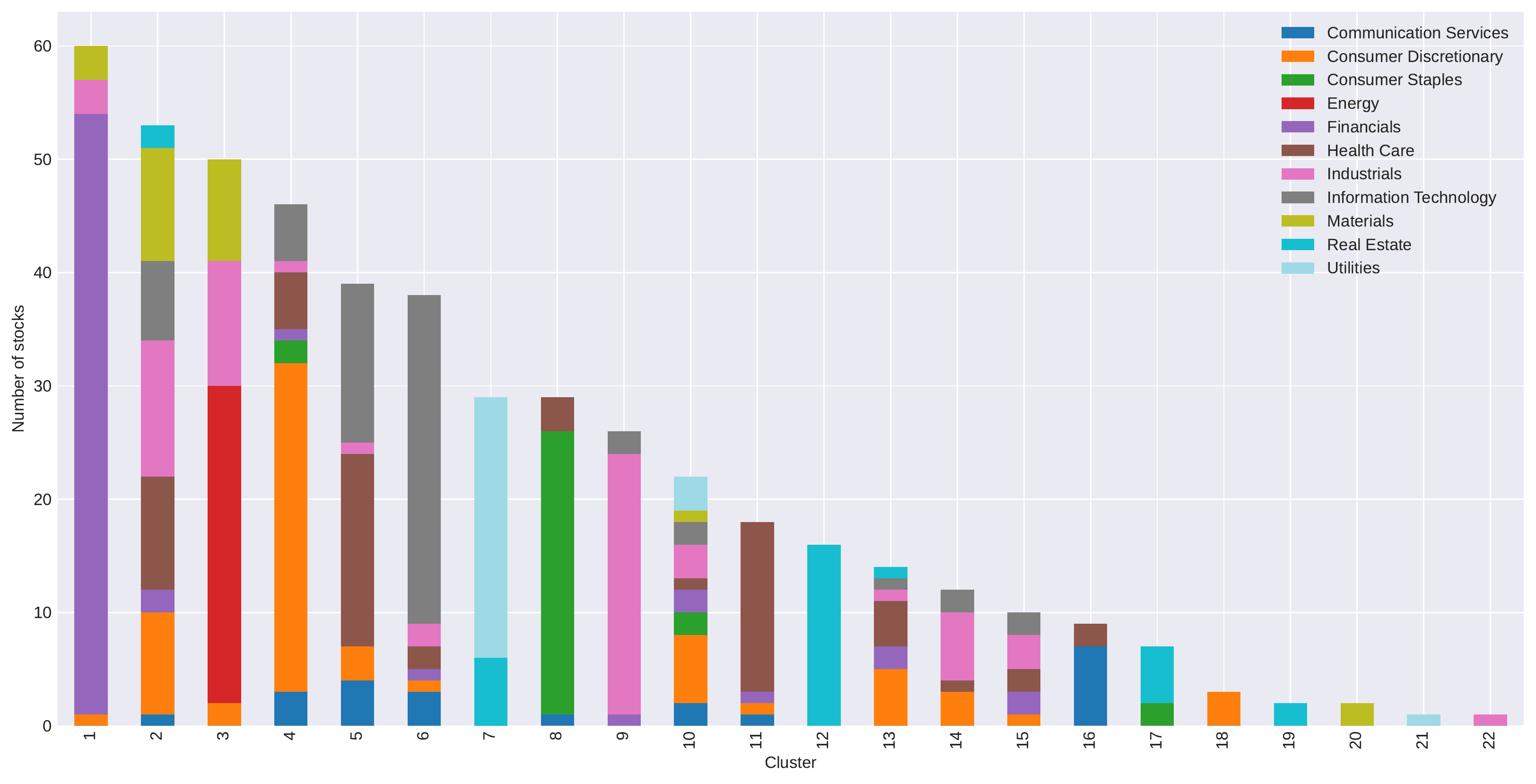}
    \caption{ACC cluster compositions on 2019-02-01 compared with GICS sectors, choosing between 15 and 25 clusters.}
    \label{fig:clustering_sector_15-25_2019-02-01}
\end{figure}

\quad To examine the compositions of the clusters, we compare the clusters with sectors defined by the Global Industry Classification Standard (GICS)\footnote{Available at \url{https://www.msci.com/gics}}.
Figure \ref{fig:clustering_sector_15-25_2019-02-01} shows the clustering result obtained on Feb $1^{st}$, 2019 from the above procedure.
We observe that some of the clusters largely overlap with GICS sectors; e.g., Cluster 9 (Industrials, specifically Aerospace \& Defense), 1 (Financials), 6 (Information Technology), 8 (Consumer Staples), 11 (Health Care), 12 (Real Estate).
However, there are also ample discrepancies between these clusters and the GICS clusters.
For instance, Cluster 6 consists of mostly stocks classified as information technology by GICS.
A few notable stocks in that cluster not classified as information technology by GICS are Amazon.com (consumer discretionary), Alphabet, Netflix, Electronic Arts (communication services), S\&P Global (Financials), and Rockwell Automation (Industrials).
Upon closer examination, one may find that those companies are highly associated with the IT industry in their business nature.
Indeed,  Amazon.com, Netflix, and Alphabet are often perceived as IT companies and usually mentioned together as members of ``FAANG''.\footnote{The FAANG includes Facebook, Amazon, Apple, Netflix, and Alphabet (Google).}
In fact, four of FAANG are in this cluster, with the exception being Facebook.
Electronic Arts, a software company that creates video games, is also naturally associated with the IT industry.
S\&P Global's primary business in financial information and analytics is likely why it is highly related to information technology, especially in today's world where finance is largely online and digital.
Although Rockwell Automation, formerly Rockwell International, was known for manufacturing aircraft and electronic components, its current business lies in control systems and software applications for industrial automation.
Another two companies in Cluster 6, yet not classified as information technology by GICS, are in the health care industry: Intuitive Surgical, which develops robotic surgical products, and Agilent Technologies, which provides analytical instruments and technology platforms for laboratories.
In these examples, our clustering appears to gather technology companies in this cluster, not according to any existing taxonomy, but by discovering the associations of their business nature through the stock market movements.
There is still, however, one exception that does not quite belong by the business nature: Fortune Brands Home \& Security, which is a manufacturer of home fixtures and hardware.

\quad The above example demonstrates a characterizing feature of the correlation blockmodel as opposed to traditional industry classifications:
the ACC algorithm, which is purely data-driven, is able to identify stock groups based on correlation similarities rather than relying on fundamental information or knowledge about the companies' business.
One advantage of this feature in asset selection and allocation is that we can uncover those ``under-the-radar'' stocks that can be used to replace the ``big name'' stocks (such as the FAANG) -- the latter often over-owned and hence tend to be over-priced -- in a well-diversified portfolio.
\begin{table}[!htb]
    \tiny
    \centering
    \begin{tabular}{|l|l|r|r|}
        \hline
        \bf{Ticker} &       \bf{Company Name} &  \bf{Ann. Sharpe Ratio} & \bf{Ann. Return} \\\hline

        FB     &                 Facebook Inc &              0.86 &      21.94\% \\\hline
        AMZN   &               Amazon.Com Inc &              1.12 &      23.62\% \\\hline
        AAPL   &                    Apple Inc &              3.81 &      89.15\% \\\hline
        NFLX   &                  Netflix Inc &              0.05 &       1.55\% \\\hline
        GOOGL  &                 Alphabet Inc &              1.24 &      28.21\% \\\hline\hline
        EA     &          Electronic Arts Inc &              0.52 &      18.39\% \\\hline
        SPGI   &               S\&P Global Inc &              2.84 &       53.3\% \\\hline
        ROK    &          Rockwell Automation &               0.6 &       16.2\% \\\hline
        ISRG   &       Intuitive Surgical Inc &              0.26 &       6.72\% \\\hline
        A      &     Agilent Technologies Inc &               0.4 &       9.53\% \\\hline
        FBHS   &  Fortune Brands Home \& Secur &              2.28 &      55.87\% \\\hline
    \end{tabular}
    \smallskip
    \caption{Performance of non-IT stocks in Cluster 6 between 2019-02-01 and 2020-02-01, compared with the FAANG stocks.}
    \label{tab:stock_perf}
\end{table}

% \begin{table}[!htb]
%     \tiny
%     \centering
%     \begin{tabular}{|l|l|r|r|}
%         \hline
%         \bf{Ticker} &       \bf{Company Name} &  \bf{Ann. Sharpe Ratio} & \bf{Ann. Return} \\\hline

%         FB     &                 Facebook Inc &              1.07 &      38.36\% \\\hline
%         AMZN   &               Amazon.Com Inc &              1.31 &      38.78\% \\\hline
%         AAPL   &                    Apple Inc &              1.96 &      69.11\% \\\hline
%         NFLX   &                  Netflix Inc &              0.47 &      18.35\% \\\hline
%         GOOGL  &                 Alphabet Inc &              1.36 &      41.97\% \\\hline\hline
%         EA     &          Electronic Arts Inc &              0.61 &       20.8\% \\\hline
%         SPGI   &               S\&P Global Inc &              1.13 &      37.78\% \\\hline
%         ROK    &          Rockwell Automation &              0.77 &      28.63\% \\\hline
%         ISRG   &       Intuitive Surgical Inc &               0.8 &      28.57\% \\\hline
%         A      &     Agilent Technologies Inc &              1.15 &      33.47\% \\\hline
%         FBHS   &  Fortune Brands Home \& Secur &              0.95 &      39.21\% \\\hline
%     \end{tabular}
%     \smallskip
%     \caption{Performance of non-IT stocks in Cluster 6 between 2019-02-01 and 2021-07-23, compared with the FAANG stocks.}
%     \label{tab:stock_perf_to_now}
% \end{table}

\quad Indeed, even if we just compare the {\it individual} performance of the six non-IT stocks in Cluster 6 with the FAANG in the one year {\it after} the clustering on February $1^{st}$, 2019, the results are noteworthy.
Table \ref{tab:stock_perf} shows the results.
In particular, Fortune Brands Home \& Security and S\&P Global have had decent annualized Sharpe ratios and returns compared to the IT stocks.
This shows the potential of utilizing the clustering information to identify less-known stocks that have good performance and provide good diversification in a portfolio.

\quad In addition to Cluster 6, which largely overlaps with a single GICS sector, we have Cluster 7, which consists of two closely related sectors: Real Estate and Utilities.
There are also clusters that do not show any apparent theme that aligns with any particular GICS sector.
For example, Clusters 2 and 10 both include stocks that are in all but a few GICS sectors.
This fact reaffirms that the clustering is providing information that cannot be reflected in the industrial classification nor by mere common knowledge or experience.

\quad Next, we compare ACC with two other clustering methods.
The first is the single-linkage clustering, which is an instance of the classical hierarchical clustering method \citep{ANDERBERG1973131}.
The same method is used in \cite{Man99} to analyze the hierarchical structure in the financial market.
The second method is the $k$-medoids method \citep{Kaufman1990}, which is based on the search of $k$ representative stocks as medoids and the assignment of every other stock to the closest medoid.
These two clustering methods are briefly reviewed in Appendices B.1 and B.2.
Note here that the two methods for asset clustering are purely heuristic, and they are both based on Criterion 1 only, namely to cluster in a way that assets in the same group have high correlations.

\begin{figure}[!htb]
    \centering
    \begin{subfigure}[b]{0.45\textwidth}
        \centering
        \includegraphics[width=\textwidth]{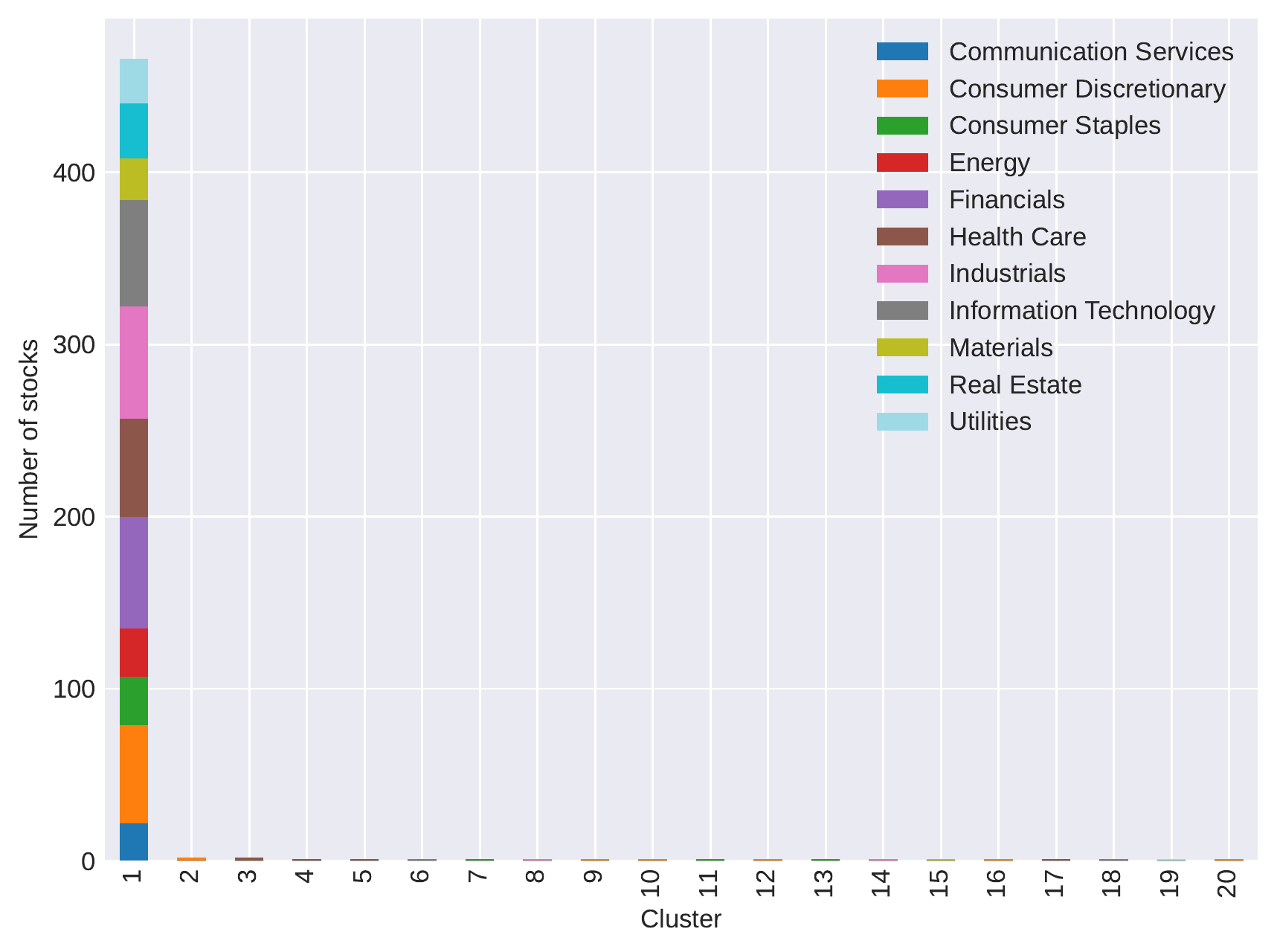}
        \caption{Hierarchical clustering}
        \label{fig:hierarchical_clustering_sector_20_2019-02-01}
    \end{subfigure}
    \begin{subfigure}[b]{0.45\textwidth}
        \centering
        \includegraphics[width=\textwidth]{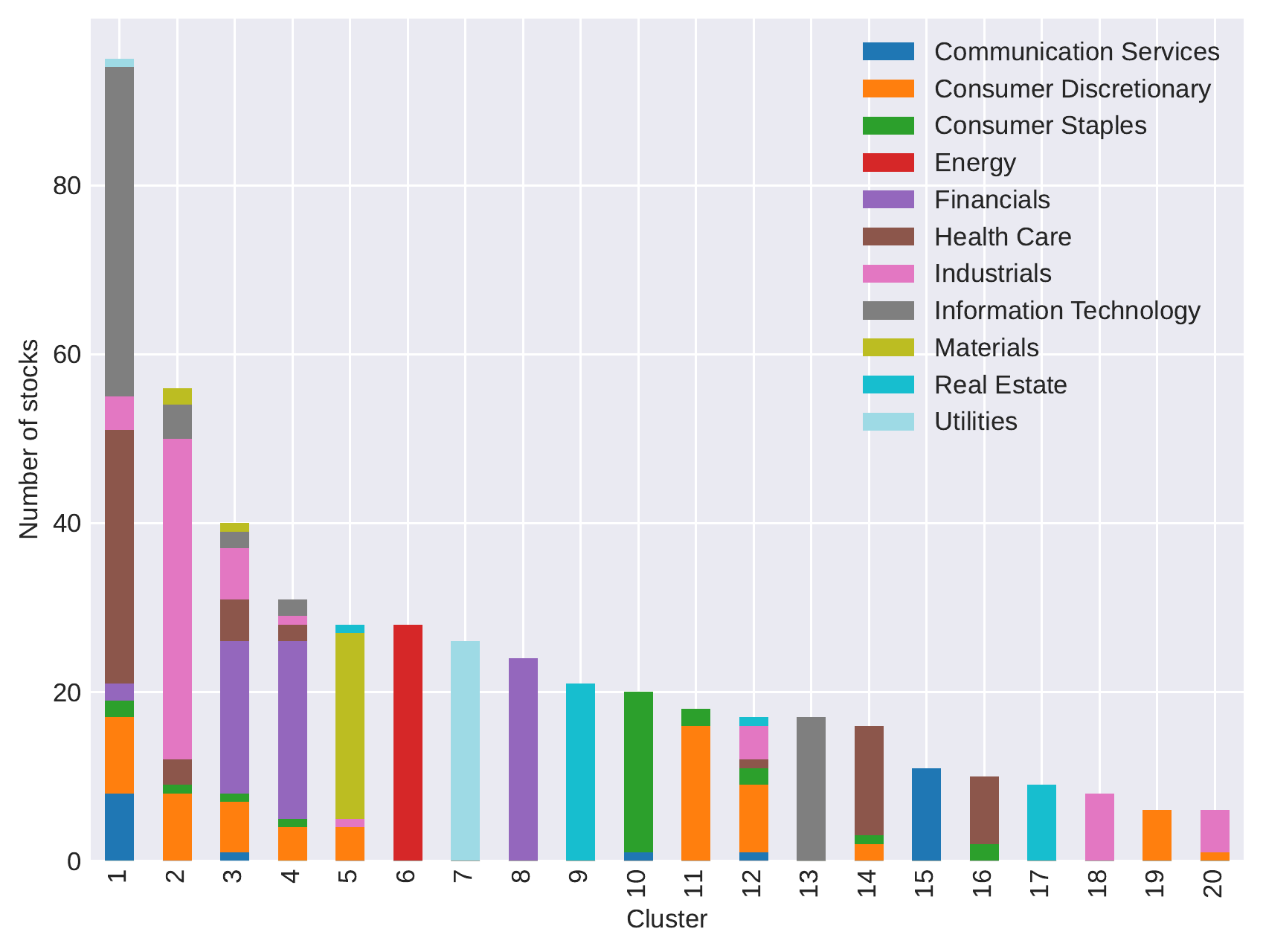}
        \caption{$k$-medoids clustering}
        \label{fig:k-medoids_clustering_sector_20_2019-02-01}
    \end{subfigure}
    \caption{Hierarchical and $k$-medoids cluster compositions  on 2019-02-01 compared with GICS sectors.}
    \label{fig:other_clustering_sector_20_2019-02-01}
\end{figure}

\quad Figure \ref{fig:other_clustering_sector_20_2019-02-01} shows their results obtained on Feb $1^{st}$, 2019.
We observe in Figure \ref{fig:hierarchical_clustering_sector_20_2019-02-01} that the clusters obtained from the hierarchical clustering are not very useful, with almost all stocks concentrating in one giant cluster, while the other clusters are mostly singletons.
This finding is robust across all sliding windows that we tested.
Results with similar characteristics are also reported in \cite{Musmeci2015a}.
These results show that hierarchical clustering is very sensitive to the existence of global factors that load on a large number of stocks.
When most or all stocks are sufficiently correlated with one another, those stocks will be merged with priority by hierarchical clustering and form an oversized cluster.

\begin{figure}[!htb]
    \centering
    \includegraphics[width=0.72 \textwidth]{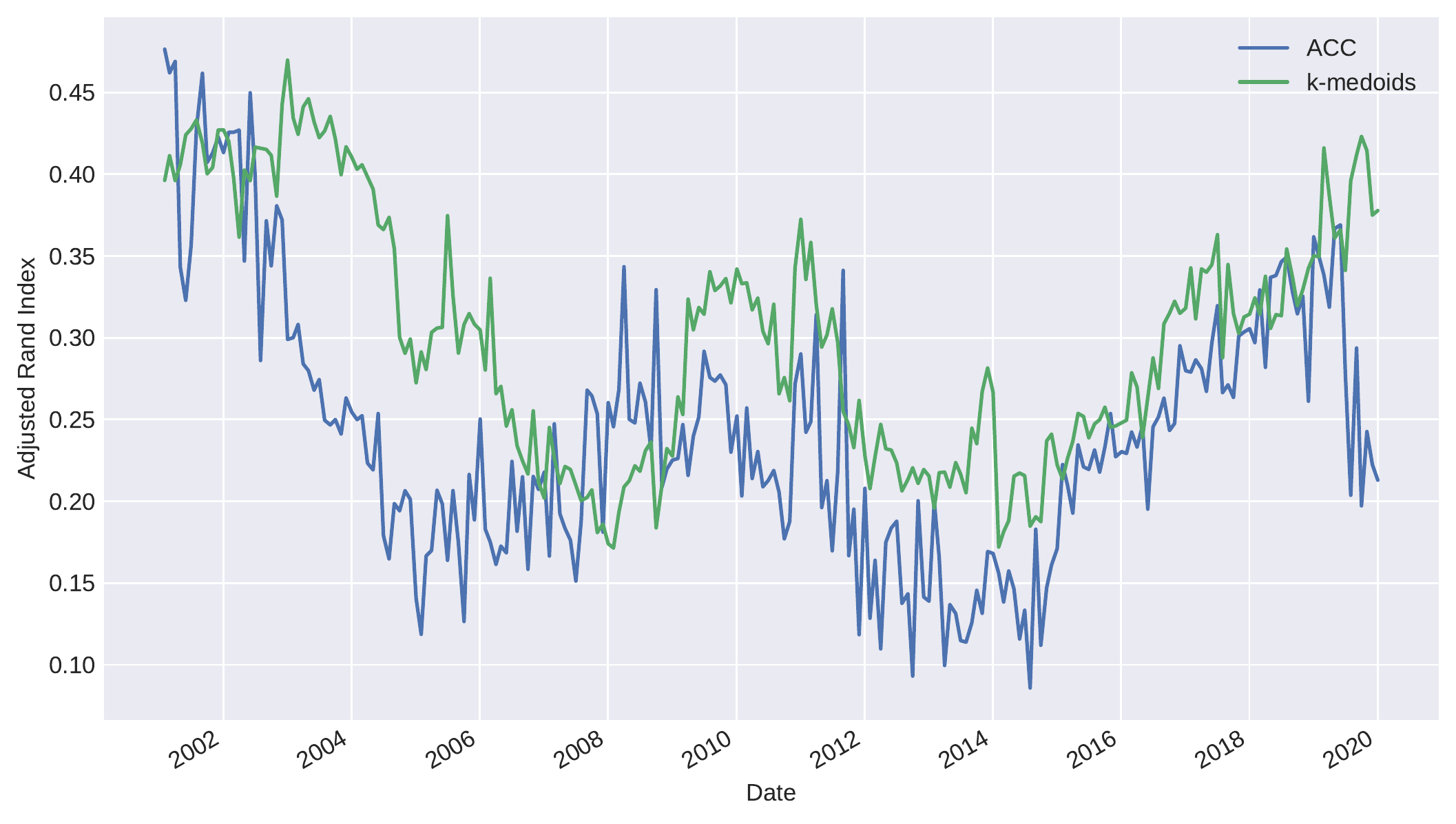}
    \caption{The adjusted Rand index $R_{adj}$ of ACC and $k$-medoids clusterings compared with GICS sectors}
    \label{fig:adj_rand_index}
\end{figure}

\quad  The clusters from $k$-medoids, as shown in Figure \ref{fig:k-medoids_clustering_sector_20_2019-02-01}, also overlap to some extent with the GICS sectors; yet the compositions are sufficiently different from our ACC results.
For example, $k$-medoids puts Financial stocks mainly into three clusters: Cluster 3, Cluster 4, and Cluster 8, in each of which Financials are the majority, whereas ACC groups most Financial stocks together in Cluster 1 of Figure \ref{fig:clustering_sector_15-25_2019-02-01}.
In the ACC partition, Consumer Discretionary stocks are more concentrated in a single cluster (Cluster 4), while in $k$-medoids clustering, they are more scattered across different clusters.
In ACC, we have observed that four of the FAANG stocks are grouped in the same cluster, together with many other IT stocks and six stocks from other sectors.
In $k$-medoids, all FAANG stocks are grouped together in Cluster 1, which is the largest cluster containing a majority of the IT stocks but also stocks in Utilities, Industrials, Health Care, Financials, Consumer Staples, Consumer Discretionary, and Communications Services.
Visually, it appears that the $k$-medoids clusters are more similar to GICS sectors than ACC.
To quantify this similarity, we calculate the adjusted Rand index $R_{adj}$ \citep{Hubert1985b} between the clustering and the GICS sectors.
$R_{adj}$ is an index that measures the similarity between two different partitions on the same set of objects, with the value $0$ representing no similarity and the value $1$ representing identical partitions.
We calculate $R_{adj}$ for both ACC and $k$-medoids clusterings compared to GICS sectors for each month.
The results are presented in Figure \ref{fig:adj_rand_index}.
Indeed, $k$-medoids clustering almost always produces more similar results to the GICS sectors than ACC does. Hence, ACC provides a more distinctive alternative clustering to the existing GICS classification. In particular,
if one of the benefits of clustering stocks, as discussed earlier, is to technically unearth less known names whose price movements are similar to those of the big names, then the one producing results less similar to the GICS sectors would be advantageous.

%--------------------------------------------------------
\subsection{Portfolio construction and backtesting}
\label{sc44}
We construct portfolios based on the ACC clustering results by first selecting the stock with the lowest volatility from each cluster.
The volatility is measured using the sample variance of daily returns in the past 500 trading days.
Theoretically, this selection is dictated by Theorem \ref{thm:global_minimum_variance}.
From a practical and empirical perspective, the reason why we choose low volatility as the criterion is twofold.
First, low volatility as a criterion does not involve the estimation of the mean returns.
ACC and the $k$-medoids algorithm tested do not cluster stocks based on their mean returns but only their correlations.
So it would be inconsistent if we selected stocks from the clusters based on return-related criteria, e.g., mean returns or Sharpe ratios.
More importantly, the estimation of mean returns is well known to be often inaccurate (the ``mean-blur" problem; see e.g.
\cite{Merton1980}), thus rendering return-related criteria unreliable of indicating future performance.
The second reason is that stocks with low volatility have been observed to outperform the benchmarks over time, which is contrary to CAPM and is documented as the ``low-risk anomaly'' (e.g., \cite{Zaremba2017}).
We only choose one stock from each cluster, so the number of stocks in the portfolio equals the number of clusters discovered by the ACC algorithm.
For comparison, we select stocks using the same criterion on the results of $k$-medoids clustering, where we select $20$ stocks from $20$ clusters, and also from GICS sectors and industry groups, where we select one stock from each sector and each industry group, respectively.
The GICS sectors and industry groups are ``point-in-time'', meaning that stocks selected on a given date are based on their GICS classifications on that date in history.
There are 10 GICS sectors before September 2016 and 11 afterward.
Similarly, there are 22 GICS industry groups before December 2001, 23 before April 2003, and 24 afterward.
These numbers determine the numbers of stocks selected in history based on the GICS sectors and industry groups.

\quad Once a set of stocks is determined by the above procedure, three asset allocation strategies are employed and compared.
The first strategy is the risk parity strategy, which, since the 2008 crisis, has become one of the most popular approaches among portfolio managers.
The second strategy is the minimum variance allocation strategy.
The third strategy is Markowitz's mean-variance strategy without short-selling, where we set the target annualized return to $10$\%.
We briefly review these three allocation strategies in Appendices B.3 and B.4.

\begin{figure}[!htb]
    \centering
    \includegraphics[width=0.7 \textwidth]{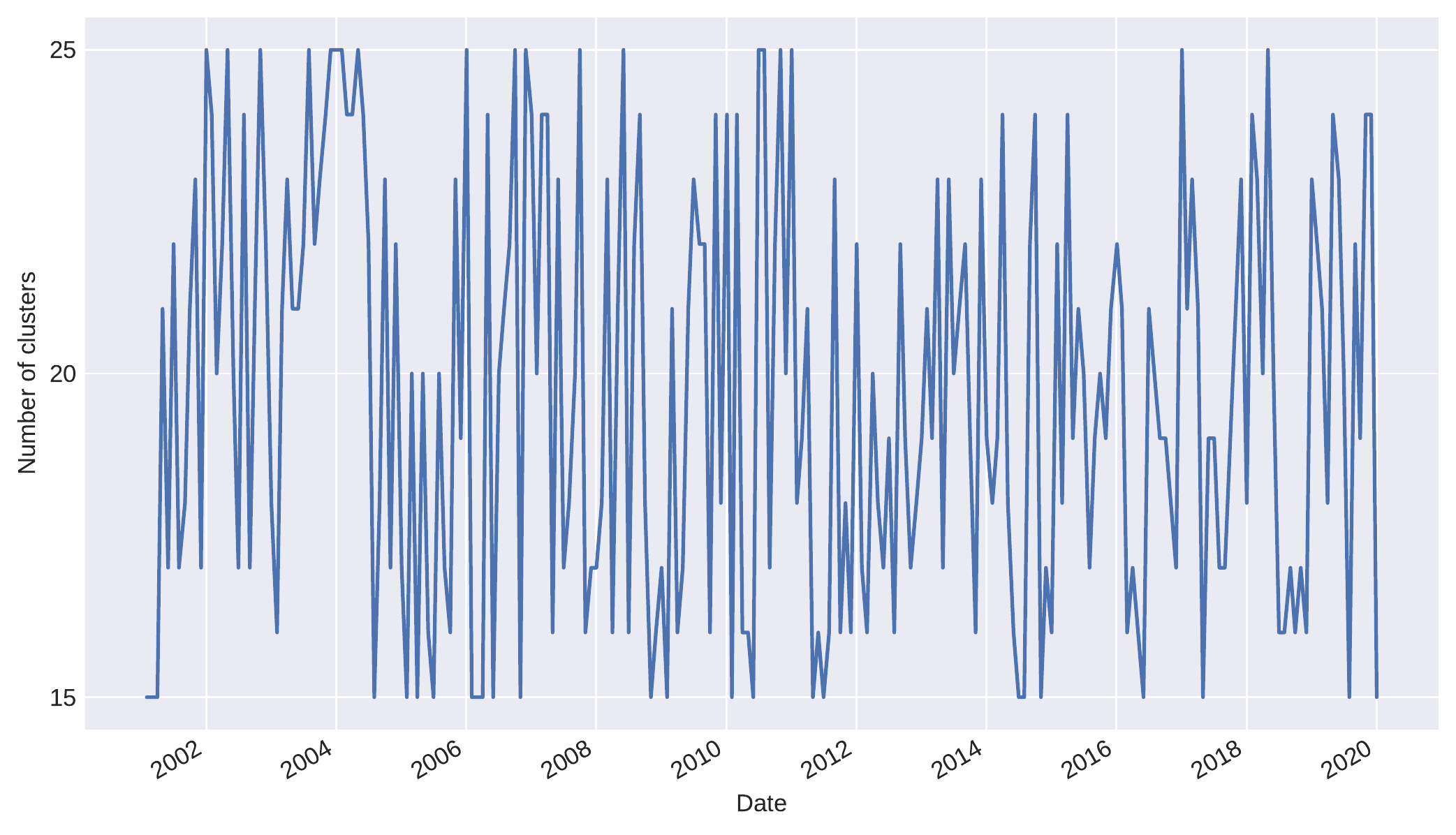}
    \caption{Number of clusters discovered by ACC over time.}
    \label{fig:nclusters_15-25}
\end{figure}

\quad Recall that for ACC, the number of the clusters in Rule \ref{rule:nb3} is set between $15$ and $25$, yielding $15$ to $25$ stocks.
The exact number of clusters discovered by ACC over time, reported in Figure \ref{fig:nclusters_15-25}, fluctuates wildly, and hence the resulting clusters are likely very different between months.
If we were to construct a portfolio using the clustering results every month, we would be in and out of positions very frequently as the stocks in the portfolio will likely be changing from month to month, which is not desirable.
Therefore, we will only readjust the portfolios once every year. Specifically, for each of the clustering methods,
on the first trading day of each February, a new set of stocks is selected according to the clustering result, and their allocations under the three aforementioned strategies are respectively calculated using all daily returns in the past 500 trading days, starting with the first day when all stocks are available.
The positions are then held until the first trading day of the following February.
Any dividends are immediately reinvested in the same stock.
We assume no transaction cost for simplicity.

\quad As benchmarks, we first take the S\&P 500 ETF (NYSE ticker: SPY), which is the largest ETF in the world and designed to track the S\&P 500 stock market index.
We also create portfolios consisting of all S\&P 500 sector ETFs\footnote{\url{https://www.ssga.com/us/en/individual/etfs/capabilities/invest-with-sector-etfs/sector-and-industry-etfs}} using the above three allocation strategies.
Each of these ETFs consists of companies in a specific GICS sector in the S\&P 500 Index.
Investing in these ETFs represents a simple method of diversifying among sectors while maintaining the ability to decide the weight of each sector optimally.

\begin{figure}[!htb]
    \centering
    \begin{subfigure}[b]{\textwidth}
        \centering
        \includegraphics[width=0.72 \textwidth]{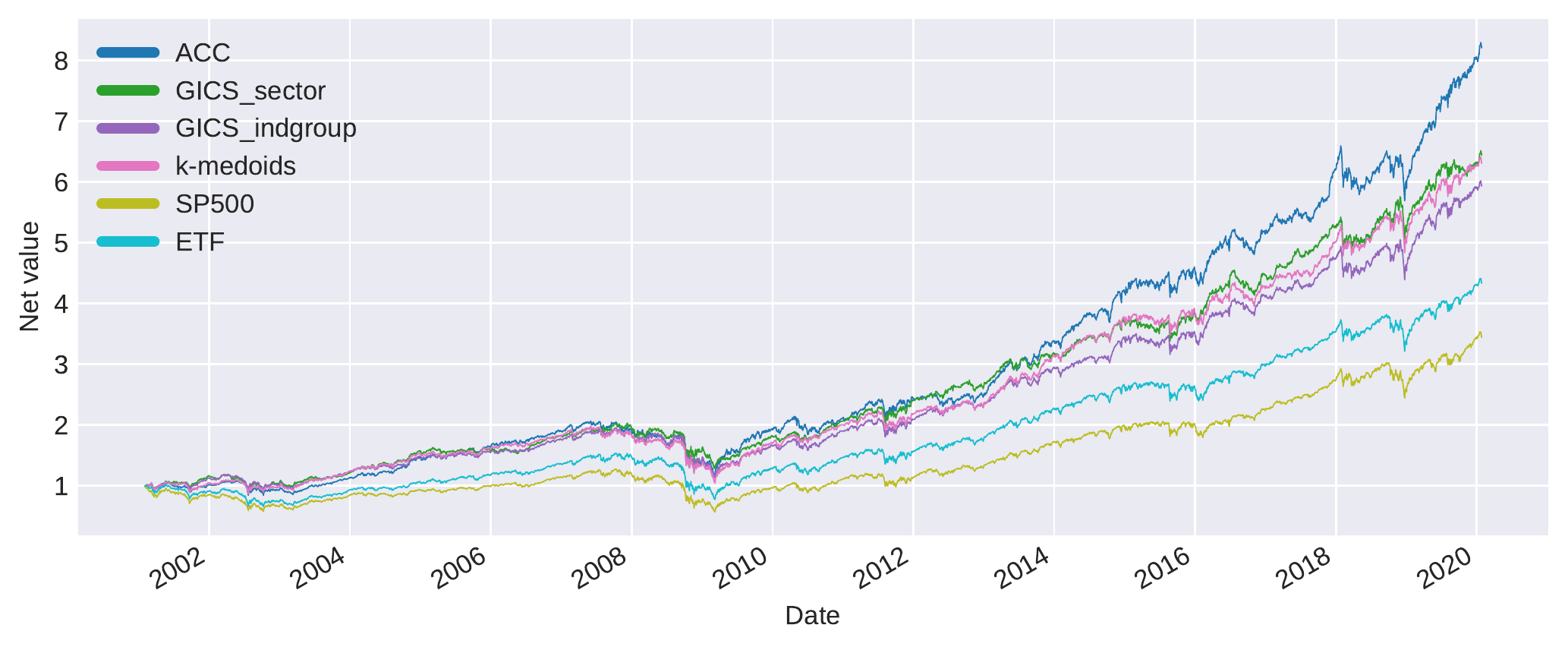}
        \caption{Risk parity portfolios}
        \label{fig:backtesting_15-25_clusters_risk_par_annually}
    \end{subfigure}
    \begin{subfigure}[b]{\textwidth}
        \centering
        \includegraphics[width=0.72 \textwidth]{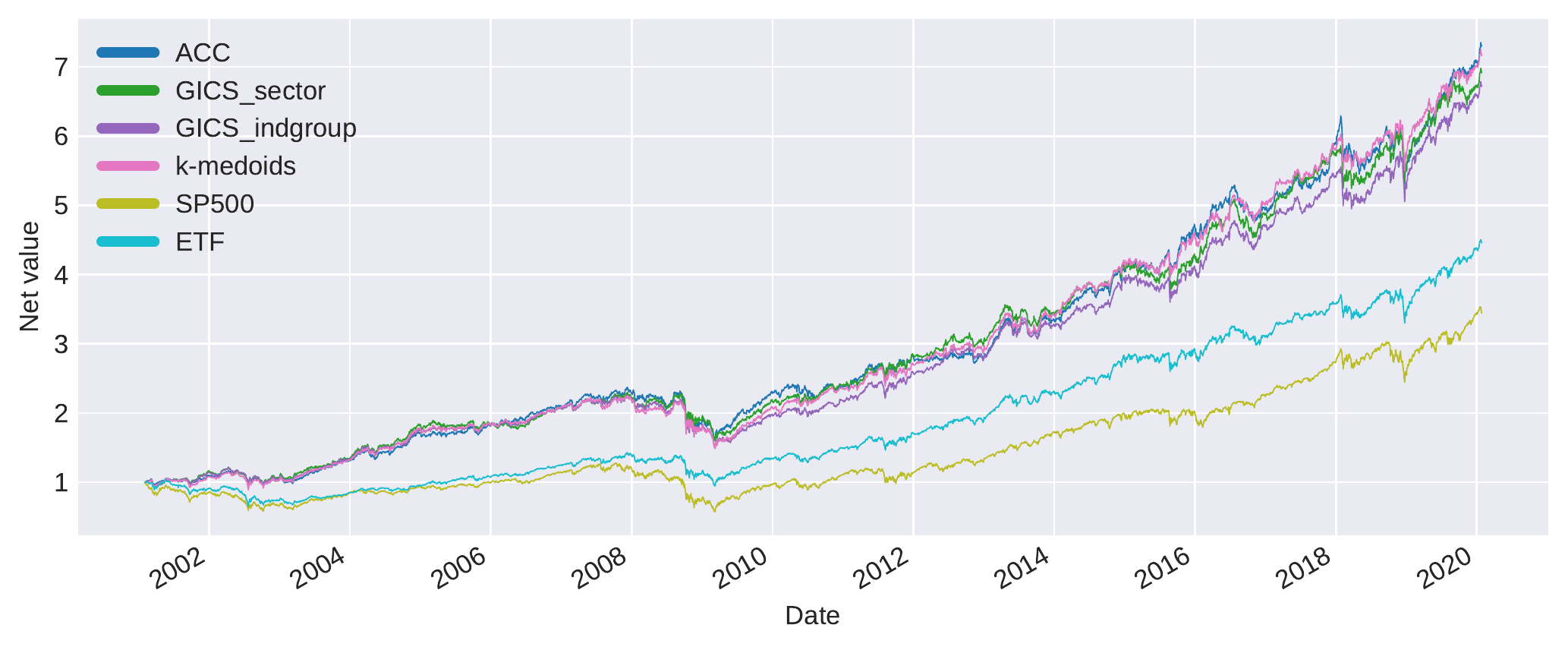}
        \caption{Minimum variance portfolios}
        \label{fig:backtesting_15-25_clusters_min_var_annually}
    \end{subfigure}
    \begin{subfigure}[b]{\textwidth}
        \centering
        \includegraphics[width=0.72 \textwidth]{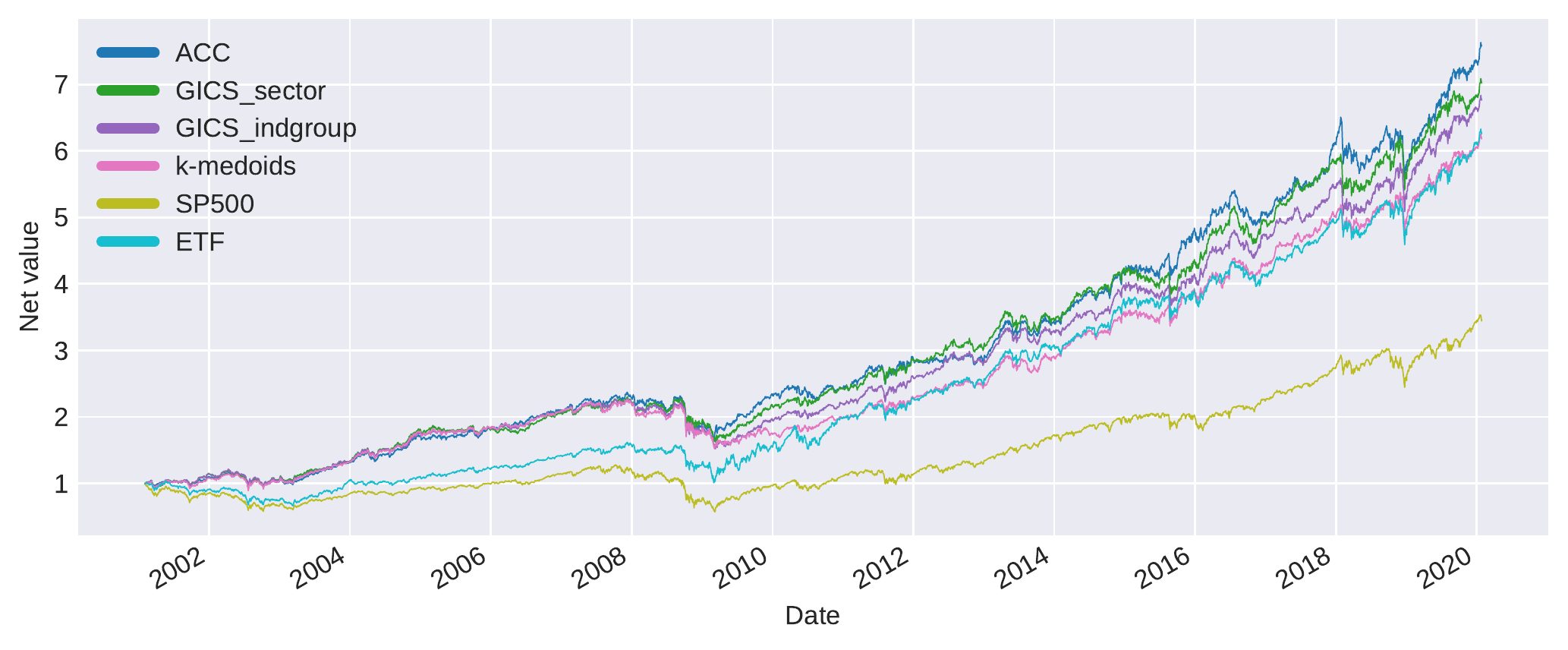}
        \caption{Single-period mean-variance portfolios with 10\% target return}
        \label{fig:backtesting_15-25_clusters_mean_var_annually}
    \end{subfigure}
    \caption{Daily value comparison among the portfolios. ACC chooses between 15 and 25 stocks, and $k$-medoids chooses 20 stocks. Rebalanced annually.}
    \label{fig:backtesting_15-25_clusters_annually}
\end{figure}

\quad Figure \ref{fig:backtesting_15-25_clusters_annually} shows the daily values of these portfolios along with the benchmark, and Tables \ref{tab:backtesting_15-25_clusters_risk_par_annually} -- \ref{tab:backtesting_15-25_clusters_mean_var_annually} report results based on performance metrics commonly used in the wealth management industry.
ACC outperforms SPY significantly in the most important return and risk metrics (including the Sharpe, Sortino, and Calmar ratios, annual volatility, and annualized return) under all three strategies.
ACC also has a much smaller maximum drawdown than SPY.
It also significantly outperforms the portfolios of sector ETFs under all allocation strategies.
This observation shows the advantage of asset selection using clustering over simply constructing portfolios using the sector ETFs.
Between ACC and the portfolios based on GICS classifications, ACC portfolios offer better return-risk ratios and faster recovery from the maximum drawdown.
Between ACC and $k$-medoids, ACC is also superior when employing the risk parity and the mean-variance allocation strategies.
With the minimum-variance strategy, ACC still offers a better annualized return, a smaller maximum drawdown, and faster recovery than $k$-medoids while maintaining a similar overall Sharpe ratio.
Moreover, as discussed earlier, ACC has the advantage of having a theoretical foundation, whereas $k$-medoids is a pure heuristic for financial time series.

\par

\begin{table}[!htb]
    \tiny
    \centering
    \begin{tabular}{|p{2.5cm}|p{1.5cm}|p{1.5cm}|p{1.5cm}|p{1.5cm}|p{1.5cm}|p{1.5cm}|}
        \hline
        {} &                     \bf{ACC} &              \bf{GICS sector} &            \bf{GICS ind. group} &                \bf{k-medoids}&                    \bf{SPY} &                      \bf{Sector ETFs} \\ \hline

        \textbf{Ending VAMI              } &                  8206.92 &                  6439.89 &                  5931.68 &                  6305.41 &                  3442.53 &                   4334.0 \\ \hline
        \textbf{Max Drawdown             } &                   44.08\% &                   36.89\% &                    42.6\% &                   47.26\% &                   55.25\% &                   49.56\% \\ \hline
        \textbf{Peak-To-Valley           } &  2007-06-01 - 2009-03-05 &  2007-12-10 - 2009-03-09 &  2007-10-12 - 2009-03-09 &  2007-06-04 - 2009-03-09 &  2007-10-09 - 2009-03-09 &  2007-10-12 - 2009-03-09 \\ \hline
        \textbf{Recovery                 } &                 261 Days &                 445 Days &                 485 Days &                 445 Days &                 774 Days &                 492 Days \\ \hline
        \textbf{Sharpe Ratio             } &                     0.79 &                     0.74 &                     0.68 &                      0.7 &                     0.36 &                     0.49 \\ \hline
        \textbf{Sortino Ratio            } &                     1.29 &                      1.2 &                     1.09 &                     1.14 &                     0.57 &                     0.77 \\ \hline
        \textbf{Calmar Ratio             } &                     0.27 &                     0.28 &                     0.23 &                     0.22 &                     0.12 &                     0.16 \\ \hline
        \textbf{Ann. Volatility          } &                   14.83\% &                   13.98\% &                   14.48\% &                   14.48\% &                   18.63\% &                   16.34\% \\ \hline
        \textbf{Ann. Downside Volatility } &                     9.1\% &                    8.63\% &                    9.04\% &                    8.95\% &                   11.83\% &                   10.39\% \\ \hline
        \textbf{Correlation              } &                      0.9 &                     0.89 &                     0.94 &                     0.93 &                      1.0 &                     0.98 \\ \hline
        \textbf{Beta                     } &                     0.72 &                     0.67 &                     0.73 &                     0.72 &                      1.0 &                     0.86 \\ \hline
        \textbf{Ann. Return        } &                   11.75\% &                   10.33\% &                    9.85\% &                   10.21\% &                    6.74\% &                    8.05\% \\ \hline
        \textbf{Ann. Turnover Ratio} &                   80.59\% &                   51.03\% &                   42.91\% &                   67.09\% &                        - &                    6.01\% \\ \hline
        \textbf{Positive Periods         } &            2631 (55.11\%) &            2554 (53.50\%) &            2568 (53.79\%) &            2575 (53.94\%) &            2600 (54.46\%) &            2626 (55.01\%) \\ \hline
        \textbf{Negative Periods         } &            2143 (44.89\%) &            2220 (46.50\%) &            2206 (46.21\%) &            2199 (46.06\%) &            2174 (45.54\%) &            2148 (44.99\%) \\ \hline
    \end{tabular}
    \smallskip
    \caption{Performance metrics of the risk parity portfolios. ACC chooses between 15 and 25 stocks, and $k$-medoids chooses 20 stocks. Rebalanced annually.}
    \label{tab:backtesting_15-25_clusters_risk_par_annually}
\end{table}
\begin{table}[!htb]
    \tiny
    \centering
    \begin{tabular}{|p{2.5cm}|p{1.5cm}|p{1.5cm}|p{1.5cm}|p{1.5cm}|p{1.5cm}|p{1.5cm}|}
        \hline
        {} &                     \bf{ACC} &              \bf{GICS sector} &            \bf{GICS ind. group} &                \bf{k-medoids}&                    \bf{SPY} &                      \bf{Sector ETFs} \\ \hline

        \textbf{Ending VAMI              } &                  7299.02 &                  6912.71 &                  6715.02 &                   7159.2 &                  3442.53 &                  4455.71 \\ \hline
        \textbf{Max Drawdown             } &                   32.45\% &                   30.56\% &                   34.04\% &                   34.06\% &                   55.25\% &                   37.65\% \\ \hline
        \textbf{Peak-To-Valley           } &  2007-12-10 - 2009-03-09 &  2007-12-10 - 2009-03-09 &  2007-12-10 - 2009-03-09 &  2007-12-10 - 2009-03-09 &  2007-10-09 - 2009-03-09 &  2001-05-21 - 2002-07-23 \\ \hline
        \textbf{Recovery                 } &                 250 Days &                 380 Days &                 516 Days &                 377 Days &                 774 Days &                 747 Days \\ \hline
        \textbf{Sharpe Ratio             } &                     0.84 &                     0.81 &                     0.82 &                     0.85 &                     0.36 &                     0.61 \\ \hline
        \textbf{Sortino Ratio            } &                     1.38 &                     1.35 &                     1.34 &                      1.4 &                     0.57 &                     0.98 \\ \hline
        \textbf{Calmar Ratio             } &                     0.34 &                     0.35 &                     0.31 &                     0.32 &                     0.12 &                     0.22 \\ \hline
        \textbf{Ann. Volatility          } &                   13.19\% &                   13.21\% &                   12.95\% &                   12.81\% &                   18.63\% &                   13.36\% \\ \hline
        \textbf{Ann. Downside Volatility } &                     8.0\% &                    7.97\% &                     7.9\% &                    7.82\% &                   11.83\% &                    8.39\% \\ \hline
        \textbf{Correlation              } &                     0.78 &                     0.77 &                      0.8 &                     0.79 &                      1.0 &                     0.85 \\ \hline
        \textbf{Beta                     } &                     0.55 &                     0.55 &                     0.56 &                     0.54 &                      1.0 &                     0.61 \\ \hline
        \textbf{Ann. Return        } &                   11.06\% &                   10.74\% &                   10.57\% &                   10.95\% &                    6.74\% &                    8.21\% \\ \hline
        \textbf{Ann. Turnover Ratio} &                   77.78\% &                   56.17\% &                   58.56\% &                   62.22\% &                        - &                   20.78\% \\ \hline
        \textbf{Positive Periods         } &            2597 (54.40\%) &            2567 (53.77\%) &            2580 (54.04\%) &            2579 (54.02\%) &            2600 (54.46\%) &            2600 (54.46\%) \\ \hline
        \textbf{Negative Periods         } &            2177 (45.60\%) &            2207 (46.23\%) &            2194 (45.96\%) &            2195 (45.98\%) &            2174 (45.54\%) &            2174 (45.54\%) \\ \hline
    \end{tabular}
    \smallskip
    \caption{Performance metrics of the minimum variance portfolios. ACC chooses between 15 and 25 stocks, and $k$-medoids chooses 20 stocks.  Rebalanced annually.}
    \label{tab:backtesting_15-25_clusters_min_var_annually}
\end{table}
\begin{table}[!htb]
    \tiny
    \centering
    \begin{tabular}{|p{2.5cm}|p{1.5cm}|p{1.5cm}|p{1.5cm}|p{1.5cm}|p{1.5cm}|p{1.5cm}|}
        \hline
        {} &                     \bf{ACC} &              \bf{GICS sector} &            \bf{GICS ind. group} &                \bf{k-medoids}&                    \bf{SPY} &                      \bf{Sector ETFs} \\ \hline

        \textbf{Ending VAMI              } &                  7575.26 &                  7019.87 &                  6762.27 &                  6180.35 &                  3442.53 &                  6257.01 \\ \hline
        \textbf{Max Drawdown             } &                   31.16\% &                   29.51\% &                   33.21\% &                   32.64\% &                   55.25\% &                   37.55\% \\ \hline
        \textbf{Peak-To-Valley           } &  2007-12-10 - 2009-03-05 &  2007-12-10 - 2009-03-11 &  2007-12-10 - 2009-03-11 &  2007-12-10 - 2009-03-05 &  2007-10-09 - 2009-03-09 &  2007-12-10 - 2009-03-02 \\ \hline
        \textbf{Recovery                 } &                 241 Days &                 374 Days &                 517 Days &                 600 Days &                 774 Days &                 215 Days \\ \hline
        \textbf{Sharpe Ratio             } &                     0.86 &                     0.82 &                     0.82 &                     0.77 &                     0.36 &                     0.64 \\ \hline
        \textbf{Sortino Ratio            } &                     1.42 &                     1.36 &                     1.34 &                     1.26 &                     0.57 &                     1.03 \\ \hline
        \textbf{Calmar Ratio             } &                     0.36 &                     0.37 &                     0.32 &                     0.31 &                     0.12 &                     0.27 \\ \hline
        \textbf{Ann. Volatility          } &                   13.18\% &                   13.27\% &                   12.99\% &                   13.13\% &                   18.63\% &                    15.8\% \\ \hline
        \textbf{Ann. Downside Volatility } &                    7.97\% &                     8.0\% &                    7.92\% &                    8.03\% &                   11.83\% &                    9.85\% \\ \hline
        \textbf{Correlation              } &                     0.76 &                     0.77 &                     0.79 &                      0.8 &                      1.0 &                     0.84 \\ \hline
        \textbf{Beta                     } &                     0.54 &                     0.55 &                     0.55 &                     0.56 &                      1.0 &                     0.72 \\ \hline
        \textbf{Ann. Return        } &                   11.28\% &                   10.83\% &                   10.62\% &                   10.09\% &                    6.74\% &                   10.16\% \\ \hline
        \textbf{Ann. Turnover Ratio} &                   78.07\% &                   59.01\% &                   60.41\% &                   65.16\% &                        - &                   51.14\% \\ \hline
        \textbf{Positive Periods         } &            2585 (54.15\%) &            2557 (53.56\%) &            2591 (54.27\%) &            2552 (53.46\%) &            2600 (54.46\%) &            2590 (54.25\%) \\ \hline
        \textbf{Negative Periods         } &            2189 (45.85\%) &            2217 (46.44\%) &            2183 (45.73\%) &            2222 (46.54\%) &            2174 (45.54\%) &            2184 (45.75\%) \\ \hline
    \end{tabular}
    \smallskip
    \caption{Performance metrics of the single period mean-variance portfolios. ACC chooses between 15 and 25 stocks, and $k$-medoids chooses 20 stocks. Rebalanced annually.}
    \label{tab:backtesting_15-25_clusters_mean_var_annually}
\end{table}

\begin{figure}[!htb]
    \centering
    \begin{subfigure}{0.8\textwidth}
        \includegraphics[width=\textwidth]{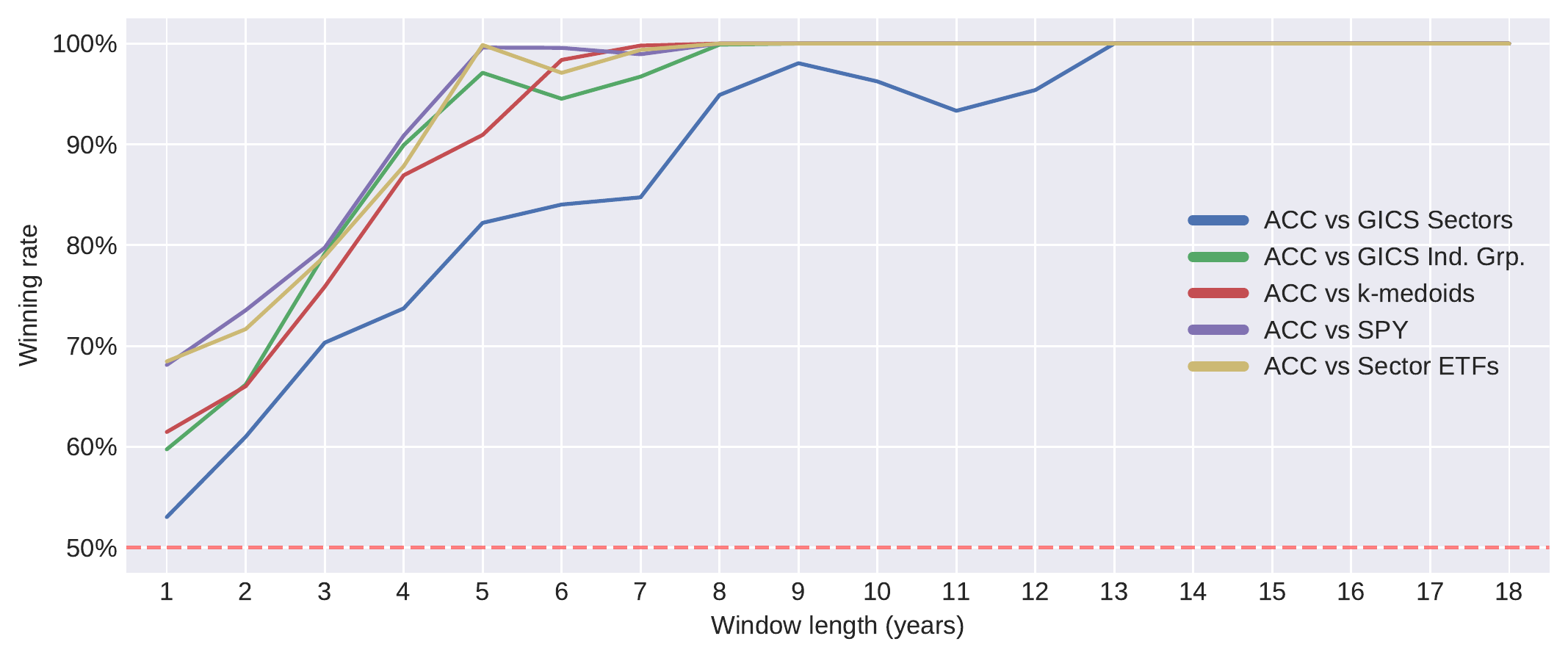}
        \caption{Risk parity}
        \label{fig:CORD_winning_rate_15-25_annually_risk_parity}
    \end{subfigure}
    \begin{subfigure}{0.8\textwidth}
        \includegraphics[width=\textwidth]{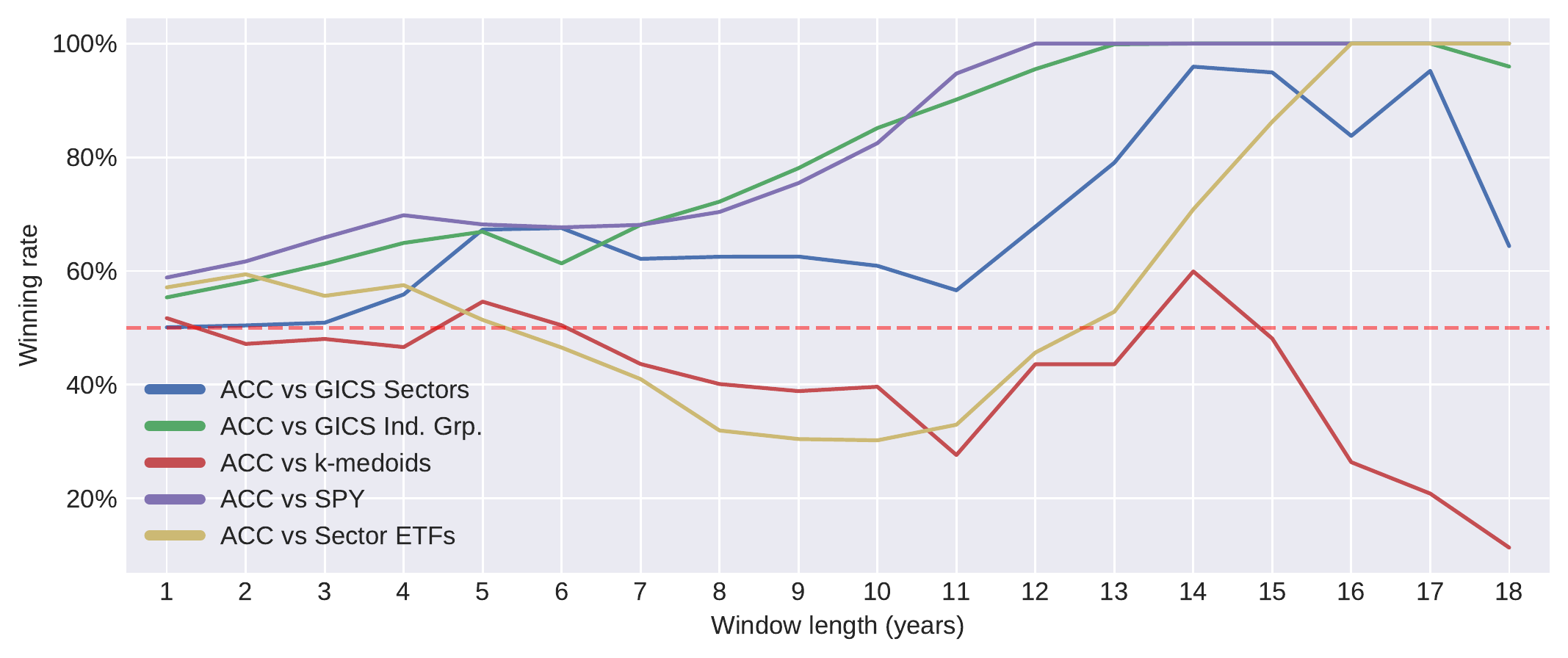}
        \caption{Minimum variance}
        \label{fig:CORD_winning_rate_15-25_annually_min_variance}
    \end{subfigure}
    \begin{subfigure}{0.8\textwidth}
        \includegraphics[width=\textwidth]{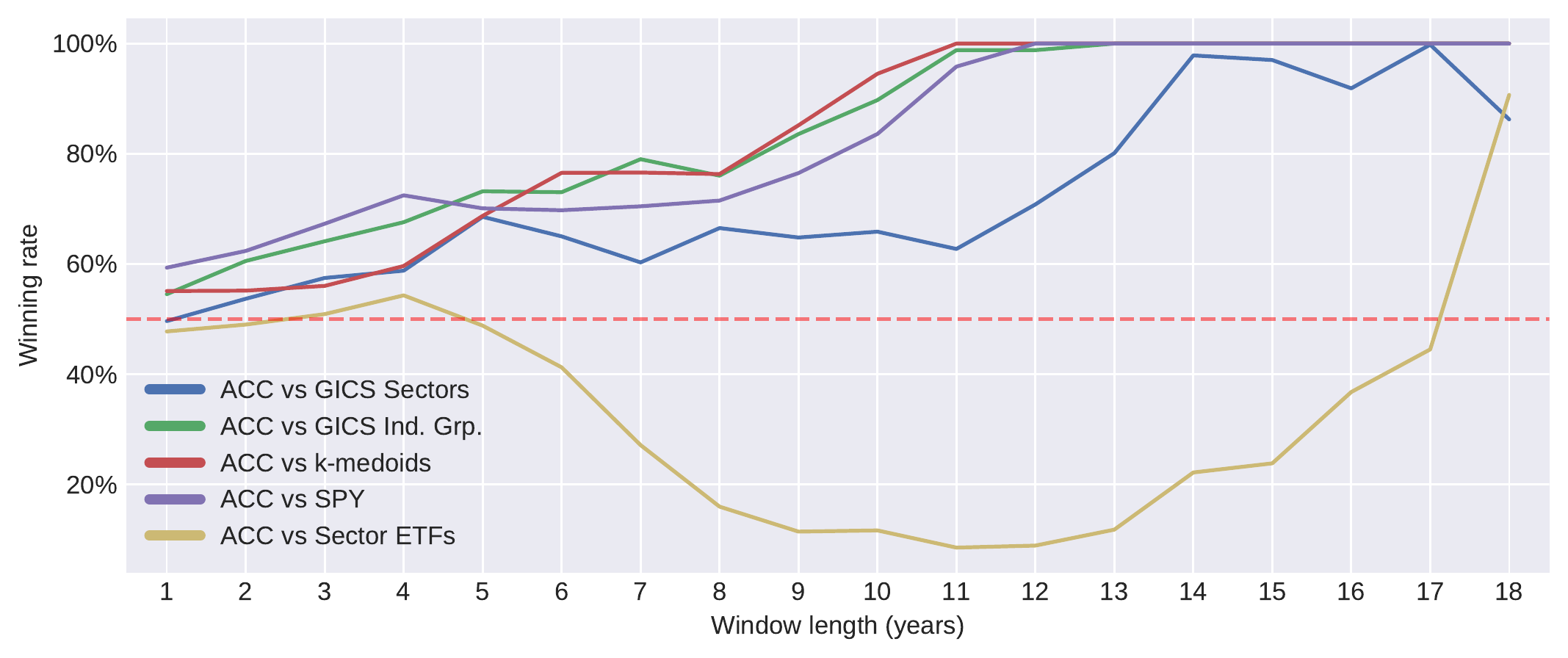}
        \caption{Mean-variance}
        \label{fig:CORD_winning_rate_15-25_annually_mean-variance}
    \end{subfigure}
    \caption{Percentage of sub-periods in which ACC has a higher return than benchmarks under different allocation strategies.}
    \label{fig:CORD_winning_rate_return}
\end{figure}

\begin{figure}[!htb]
    \centering
    \begin{subfigure}{0.8\textwidth}
        \includegraphics[width=\textwidth]{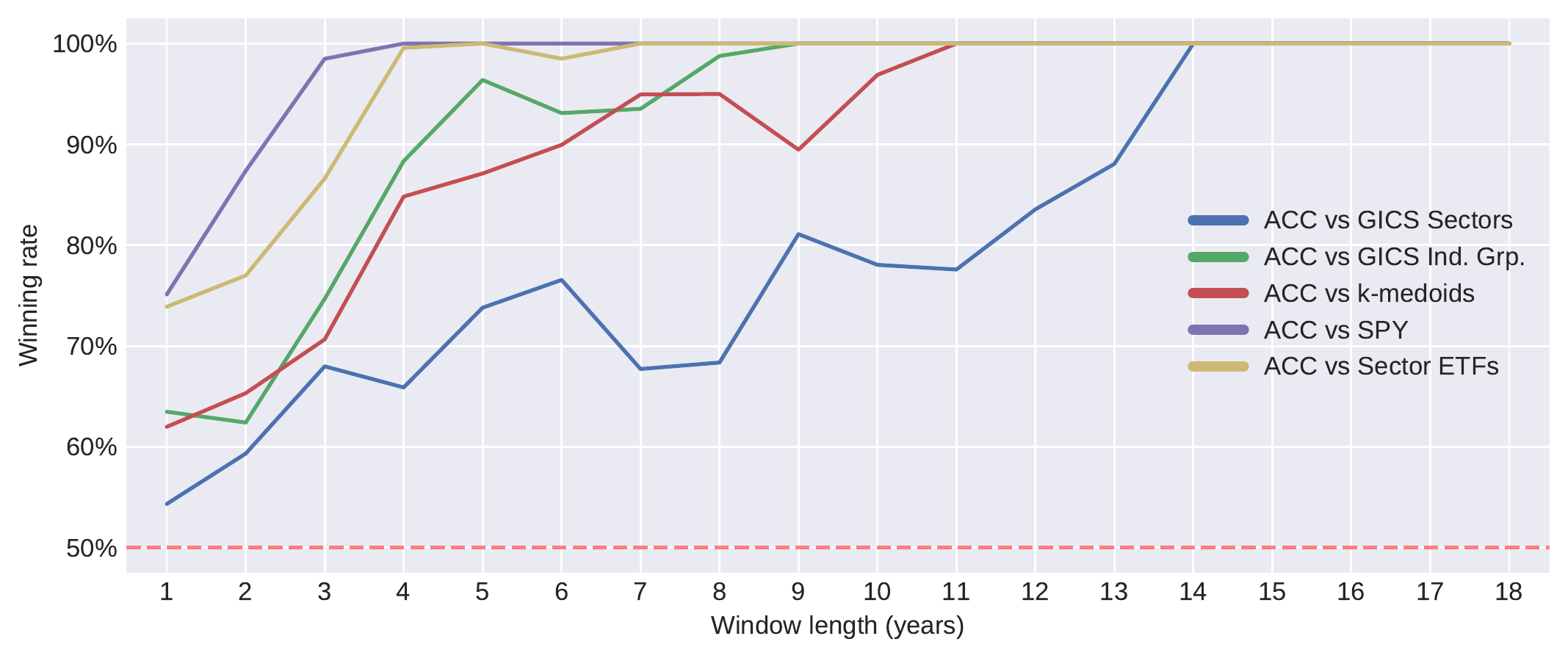}
        \caption{Risk parity}
        \label{fig:CORD_winning_rate_Sharpe_15-25_annually_risk_parity}
    \end{subfigure}
    \begin{subfigure}{0.8\textwidth}
        \includegraphics[width=\textwidth]{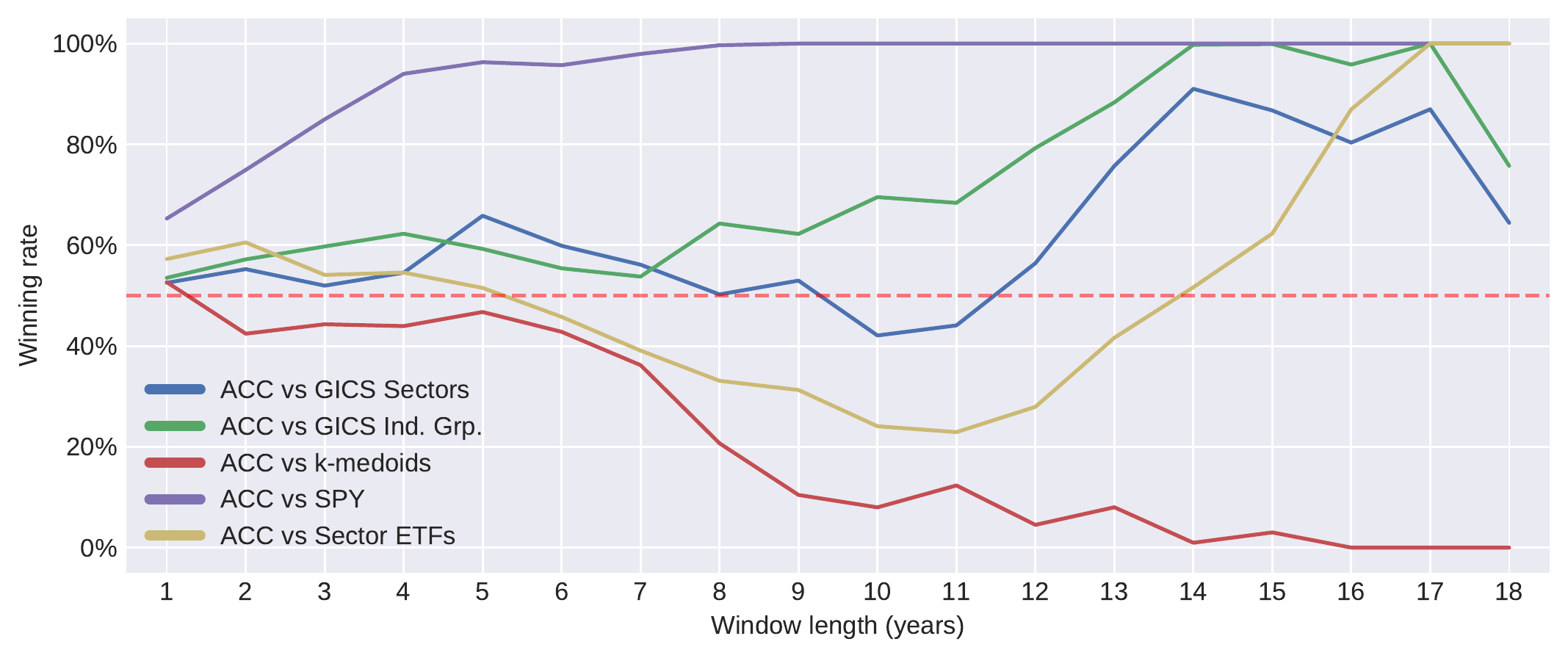}
        \caption{Minimum variance}
        \label{fig:CORD_winning_rate_Sharpe_15-25_annually_min_variance}
    \end{subfigure}
    \begin{subfigure}{0.8\textwidth}
        \includegraphics[width=\textwidth]{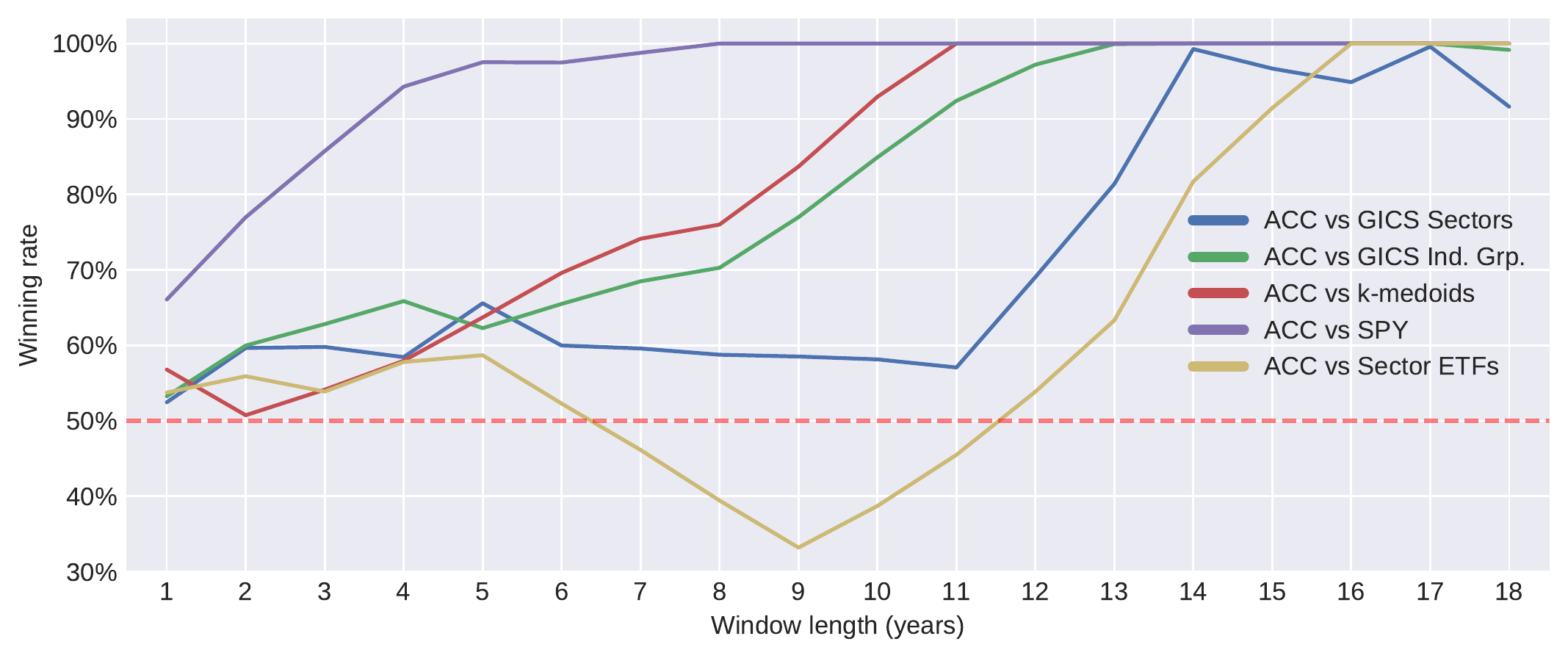}
        \caption{Mean-variance}
        \label{fig:CORD_winning_rate_Sharpe_15-25_annually_mean-variance}
    \end{subfigure}
    \caption{Percentage of sub-periods in which ACC has a higher Sharpe ratio than benchmarks under different allocation strategies.}
    \label{fig:CORD_winning_rate_Sharpe}
\end{figure}

\quad The above results compare the overall performance of the portfolios throughout the entire 19-year period.
For a more comprehensive comparison, we break down the 19-year periods into small sub-periods using a rolling window approach and compare the returns among all sub-periods.
We first set the length of the rolling window to be one calendar year and capture the returns of the portfolios within all 1-year sub-periods, e.g., from 2001-02-01 to 2002-02-01, from 2001-02-02 to 2002-02-02, and so on.
In each sub-period, we compare the annualized return and the annualized Sharpe ratio between the ACC portfolios and the benchmarks.
Then we repeat the analysis for different lengths of the rolling window.
Figure \ref{fig:CORD_winning_rate_return} shows the percentage of sub-periods of different lengths where ACC has a higher return than the benchmarks, with the three allocation strategies, respectively.
Figure \ref{fig:CORD_winning_rate_Sharpe} reports the same comparison for Sharpe ratios.\footnote{As comparing negative Sharpe ratios is meaningless, we exclude windows in which both ACC and the benchmark have negative Sharpe ratios.}

\quad As the window length approaches ten years and longer, all three portfolios from ACC clustering almost {\it always} outperform SPY, in both return and Sharpe ratio.
This observation shows that the ACC portfolios are very suitable for investors with long investment horizons.
Even for investors with short investment horizons like 1 to 2 years, ACC is still more likely to achieve better returns and Sharpe ratios than SPY.
Compared with the other benchmarks, ACC is superior in the long run to all but the $k$-medoids portfolio with the minimum variance allocation strategy.
ACC seems to underperform the portfolio of sector ETFs in annualized returns for a large range of window lengths when employing the mean-variance strategy.
Still, ACC's Sharpe ratio tends to be superior in the long run compared to the sector ETF portfolio.

\quad In addition to rebalancing the portfolios once every year (i.e., annual rebalancing), we also test semi-annual and quarterly rebalancing (i.e., rebalancing every 6 and 3 months, respectively).
Table \ref{tab:Sharpe_comparison_by_rebalance_freq} reports the results.
ACC appears to lose some advantage when the portfolios are rebalanced more frequently.
This is consistent with the previous observation that ACC clustering changes frequently, and hence, intuitively, portfolios updated frequently based on the ACC clusters might not perform well.
Still, ACC portfolios consistently outperform the SPY and the portfolios of sector ETFs.
Furthermore, with annual rebalancing, ACC not only achieves the highest Sharpe ratio with the risk parity and the mean-variance strategies among all annually rebalanced portfolios, but its Sharpe ratios are also the highest among all portfolios with the same allocation strategies regardless of rebalancing frequencies.
With the minimum variance allocation, ACC with annual rebalancing ranks third among all minimum variance portfolios with different rebalancing frequencies and only slightly lags behind the top two.
Overall, we can conclude that ACC is a consistent and robust performer with different rebalancing frequencies and strategies, and its performance stands out for slow portfolios that do not require frequent rebalancing.
It is particularly suitable for investors/funds whose investment philosophy is for less trading, if not completely ``buy and hold''.
This characteristic of ACC has practical significance, since rebalancing more frequently than annually has unfavorable tax implications and indeed investment experts have found no significant advantage of rebalancing portfolios more frequently once transaction costs and taxes are taken into consideration \citep{Zilbering2015, McNamee2019}.

\begin{table}
    \begin{subtable}{\textwidth}
        \centering
        \tiny
        \caption{Annual rebalancing}
        \begin{tabular}{|l|l|l|l|l|l|l|}
            \hline
            {} &  \bf{ACC} & \bf{GICS Sector} & \bf{GICS Ind. Grp.} & \bf{$k$-medoids} &   \bf{SPY} &   \bf{Sector ETF} \\ \hline
            \textbf{Risk parity} &  \bf{0.79} &        0.74 &          0.68 &       0.7 &  0.36 &  0.49 \\ \hline
            \textbf{Min. variance } &  0.84 &        0.81 &          0.82 &      \bf{0.85} &  0.36 &  0.61 \\ \hline
            \textbf{Mean-variance} &  \bf{0.86} &        0.82 &          0.82 &      0.77 &  0.36 &  0.64 \\ \hline
        \end{tabular}
    \end{subtable}
    \begin{subtable}{\textwidth}
        \centering
        \tiny
        \caption{Semi-annual rebalancing}
        \begin{tabular}{|l|l|l|l|l|l|l|}
            \hline
            {} &  \bf{ACC} & \bf{GICS Sector} & \bf{GICS Ind. Grp.} & \bf{$k$-medoids} &   \bf{SPY} &   \bf{Sector ETF} \\ \hline
            \textbf{Risk parity  } &  0.72 &       \bf{0.79} &          0.69 &      0.66 &  0.36 &  0.48 \\ \hline
            \textbf{Min. variance } &  0.79 &        \bf{0.86} &          0.82 &      0.75 &  0.36 &  0.59 \\ \hline
            \textbf{Mean-variance} &  0.78 &        \bf{0.83} &          0.81 &      0.71 &  0.36 &  0.56 \\ \hline
        \end{tabular}
    \end{subtable}
    \begin{subtable}{\textwidth}
        \centering
        \tiny
        \caption{Quarterly rebalancing}
        \begin{tabular}{|l|l|l|l|l|l|l|}
            \hline
            {} &  \bf{ACC} & \bf{GICS Sector} & \bf{GICS Ind. Grp.} & \bf{$k$-medoids} &   \bf{SPY} &   \bf{Sector ETF} \\ \hline
            \textbf{Risk parity  } &  0.72 &        \bf{0.76} &          0.69 &      0.65 &  0.36 &  0.48 \\ \hline
            \textbf{Min. variance } &  0.81 &        \bf{0.83} &          0.82 &      \bf{0.83} &  0.36 &  0.59 \\ \hline
            \textbf{Mean-variance} &  0.78 &        0.78 &          \bf{0.81} &       0.8 &  0.36 &   0.5 \\ \hline
        \end{tabular}
    \end{subtable}
    \caption{Sharpe ratio comparisons with different rebalancing frequencies} \label{tab:Sharpe_comparison_by_rebalance_freq}
\end{table}

\begin{figure}[!htb]
    \includegraphics[width=0.72\textwidth]{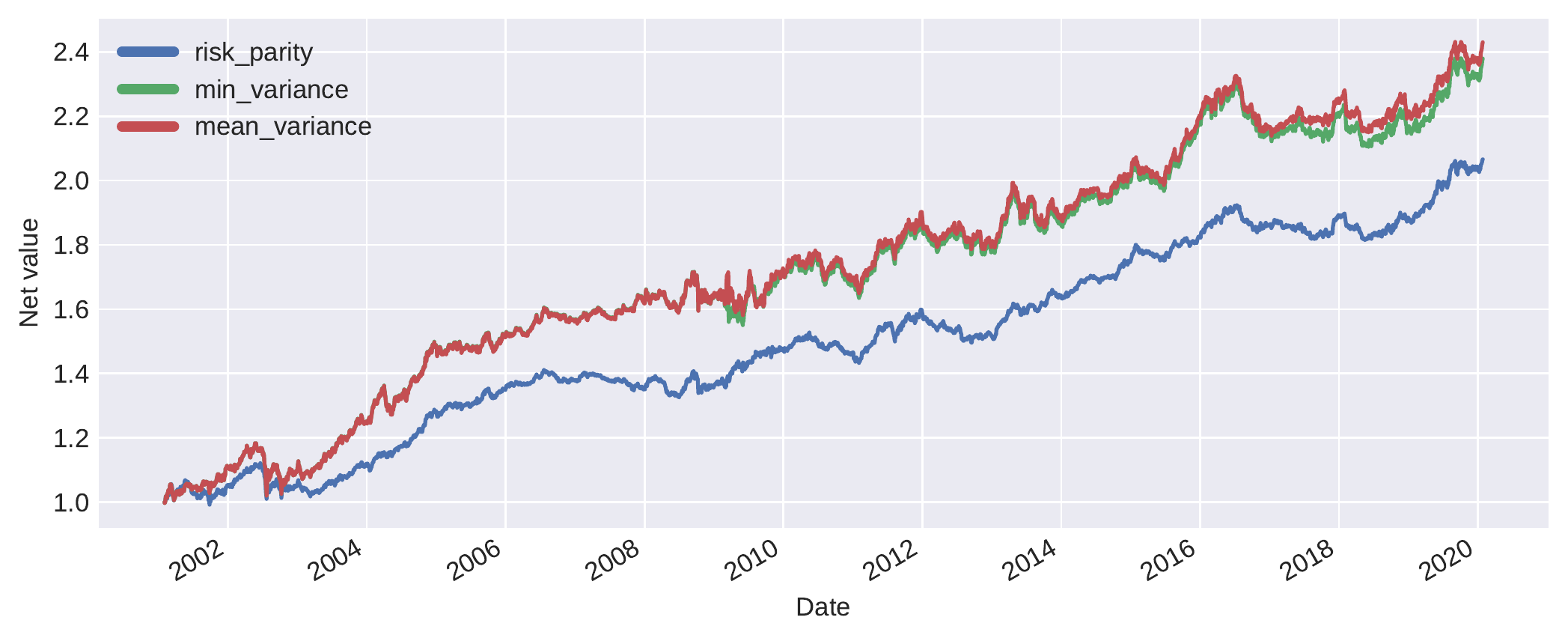}
    \caption{Performance of ACC market neutral portfolios}
    \label{fig:backtesting_CORD_market_neutral}
\end{figure}

\quad Seeing that the ACC portfolio consistently outperforms the benchmark SPY, it is intriguing to test a market-neutral portfolio that only captures the difference between the ACC portfolio and SPY.
We construct a market-neutral portfolio using each of the annually rebalanced ACC portfolios that employ the three different allocation strategies.
At each rebalancing point (the end of the first trading day of each February) and for each stock $i$ in the portfolio, we calculate its beta by linear regression of its return against the market return:\[
    X_i = \alpha_i + \beta_i R_m + \varepsilon.
\]
The beta has a closed-form solution, \[\beta_i=\frac{\cov(X_i, R_m)}{\var(R_m)},\]
and is estimated using the sample covariance between stock returns and SPY returns and the sample variance of SPY returns, both of the past 500 trading days.
Then, given a set of weights $\mathbf{w} = (w_1, w_2, \ldots, w_d)$ determined by one of the three allocation strategies, we calculate the total beta of the portfolio: \[
    \beta = \sum_{i}^d w_i\beta_i.
\]
Then, if we add to the portfolio a short position on SPY of weight $-\beta$, the portfolio will have zero exposure to SPY and hence become market neutral.
However, now the portfolio has a leverage ratio of $1+\beta$, so we scale the position down by dividing all positions by $1+\beta$.
The final weights of the market neutral portfolio are \[
    \left(\frac{w_1}{1+\beta}, \ldots,  \frac{w_d}{1+\beta},  -\frac{\beta}{1+\beta}\right),
\]corresponding to the $d$ stocks in the portfolio and the SPY.
Figure \ref{fig:backtesting_CORD_market_neutral} and Table \ref{tab:backtesting_CORD_market_neutral} show the net values and the performance metrics of the market neutral portfolios based on the ACC clusters.
Though the market-neutral portfolios have lower annualized returns, the risk parity market-neutral portfolio has a much higher Sharpe ratio than the original risk parity portfolio.
This provides an alternative application of the ACC clustering results.
If we zoom in on the period of the financial crisis (Dec 2007 -- Jul 2009), we can see in Table \ref{tab:backtesting_CORD_market_neutral_financial_crisis} that the market-neutral portfolios are very resilient to market downturns. All three portfolios obtain positive returns and have much lower drawdowns than the benchmark SPY.
\begin{table}[!htb]
    \tiny
    \centering
    \begin{tabular}{|p{4cm}|p{3cm}|p{3cm}|p{3cm}|}
        \hline
        {} &             \bf{Risk parity} &             \bf{Min. variance} &            \bf{Mean-variance} \\ \hline
        \textbf{Ending VAMI              } &                  2066.03 &                  2379.17 &                  2429.93 \\ \hline
        \textbf{Max Drawdown             } &                    9.91\% &                   13.78\% &                   13.79\% \\ \hline
        \textbf{Peak-To-Valley           } &  2002-06-20 - 2002-07-23 &  2002-05-23 - 2002-07-23 &  2002-05-23 - 2002-07-23 \\ \hline
        \textbf{Recovery                 } &                 345 Days &                 254 Days &                 257 Days \\ \hline
        \textbf{Sharpe Ratio             } &                     0.93 &                     0.77 &                     0.77 \\ \hline
        \textbf{Sortino Ratio            } &                     1.58 &                     1.27 &                     1.27 \\ \hline
        \textbf{Calmar Ratio             } &                     0.39 &                     0.34 &                     0.35 \\ \hline
        \textbf{Ann. Volatility          } &                    4.21\% &                    6.05\% &                    6.26\% \\ \hline
        \textbf{Ann. Downside Volatility } &                    2.47\% &                    3.68\% &                    3.78\% \\ \hline
        \textbf{Correlation              } &                     0.73 &                     0.97 &                      1.0 \\ \hline
        \textbf{Beta                     } &                     0.49 &                     0.94 &                      1.0 \\ \hline
        \textbf{Annualized Return        } &                     3.9\% &                    4.68\% &                     4.8\% \\ \hline
        \textbf{Annualized Turnover Ratio} &                   50.91\% &                   54.42\% &                   54.62\% \\ \hline
        \textbf{Positive Periods         } &            2522 (52.83\%) &            2508 (52.53\%) &            2505 (52.47\%) \\ \hline
        \textbf{Negative Periods         } &            2252 (47.17\%) &            2266 (47.47\%) &            2269 (47.53\%) \\ \hline
    \end{tabular}
    \smallskip
    \caption{Performance of ACC market neutral portfolios.}
    \label{tab:backtesting_CORD_market_neutral}
\end{table}

\begin{table}[!htb]
    \tiny
    \centering
    \begin{tabular}{|p{4cm}|p{2cm}|p{2cm}|p{2.5cm}|p{2cm}|}
        \hline
        {} &             \bf{Risk parity} &             \bf{Min. variance} &            \bf{Mean-variance}  &  \bf{SPY}\\ \hline
        \textbf{Ending VAMI              } &                  1056.22 &                  1015.46 &                  1023.28 &                   652.15 \\ \hline
        \textbf{Max Drawdown             } &                    4.85\% &                    9.68\% &                    7.81\% &                   53.96\% \\ \hline
        \textbf{Peak-To-Valley           } &  2008-09-15 - 2008-10-10 &  2008-09-18 - 2009-06-01 &  2008-09-18 - 2009-06-01 &  2007-12-10 - 2009-03-09 \\ \hline
        \textbf{Recovery                 } &                 123 Days &                        - &                        - &                        - \\ \hline
        \textbf{Sharpe Ratio             } &                     0.67 &                     0.11 &                     0.14 &                    -0.62 \\ \hline
        \textbf{Sortino Ratio            } &                     1.17 &                     0.18 &                     0.24 &                     -1.0 \\ \hline
        \textbf{Calmar Ratio             } &                     0.73 &                      0.1 &                     0.19 &                    -0.44 \\ \hline
        \textbf{Ann. Volatility          } &                    5.25\% &                    9.13\% &                   10.38\% &                   38.22\% \\ \hline
        \textbf{Ann. Downside Volatility } &                    3.03\% &                    5.52\% &                    6.24\% &                    23.8\% \\ \hline
        \textbf{Correlation              } &                     0.27 &                     0.27 &                     0.14 &                      1.0 \\ \hline
        \textbf{Beta                     } &                     0.04 &                     0.06 &                     0.04 &                      1.0 \\ \hline
        \textbf{Annualized Return        } &                    3.53\% &                    0.98\% &                    1.47\% &                  -23.77\% \\ \hline
        \textbf{Annualized Turnover Ratio} &                   49.69\% &                   60.41\% &                   60.15\% &                        - \\ \hline
        \textbf{Positive Periods         } &             197 (49.62\%) &             188 (47.36\%) &             185 (46.60\%) &             204 (51.39\%) \\ \hline
        \textbf{Negative Periods         } &             200 (50.38\%) &             209 (52.64\%) &             212 (53.40\%) &             193 (48.61\%) \\ \hline
    \end{tabular}
    \smallskip
    \caption{Performance of ACC market neutral portfolios between December 2007 and July 2009.}
    \label{tab:backtesting_CORD_market_neutral_financial_crisis}
\end{table}

%-------------------------------------------------------------------------------------------------
\section{Conclusion}
\label{sc5}

\quad This paper aims to identify a smaller set of stocks that attains an adequate level of diversification compared to the whole universe of stocks.
We achieve this by clustering financial assets via exploring the correlation structure.
We cluster the assets in a group according to the joint correlation with all other assets.
The idea is formalized by the correlation blockmodel, and the ACC algorithm is devised to cluster the model.
We provide rigorous analysis of the ACC algorithm and give practical guidance based on the theoretical results.
Numerical experiments show that portfolios constructed based on the ACC clustering algorithm achieve good performance compared to the market benchmark and also other portfolios consisting of similar numbers of assets.

\quad Our work can be extended in several directions.
One is to further improve the ACC algorithm.
For instance, how to recover the clusters of the correlation blockmodel when the minimal separation
$\min_{i \stackrel{G^{\star}}{\nsim} j} \cord(i,j)$ is below $\max(\sqrt{\log d \,/n}, (\log d)^{2/\alpha}/n)$? The other
is to make use of the past return information to embed Criteria  \ref{criterion:nb1} \& \ref{criterion:nb2} directly into  mean-variance optimization algorithms.
\newpage

%-------------------------------------------------------------------------------------------------
\section*{Appendix A. Proofs of theoretical results}
\label{sc3}

\quad In this section, we prove the main results -- Theorems \ref{thm:uniquepart} -- \ref{thm:complexity}.

%--------------------------------------------------------
\subsection*{A.1. Proof of Theorem \ref{thm:uniquepart}}
\label{sc31}
Assume that $\pmb{\rho} = \pmb{Z} \pmb{\Pi} \pmb{Z}^{\top} + \pmb{\Gamma}$ holds for a membership matrix $\pmb{Z}$ associated with some partition $G = \{G_1, G_2, \ldots\}$.
Then
\begin{equation*}
\max_{l \ne i, j} |\rho_{il} - \rho_{jl}| = 0 \quad \mbox{for any } i,j \in G_k.
\end{equation*}
So each group $G_k$ of $G$ is included in one of the equivalence classes $\stackrel{G^{\star}}{\sim}$ defined by \eqref{eq:eqvclass}.
As a result, the partition $G$ is finer than $G^{\star}$.
This implies that $G^{\star}$ is the unique coarsest partition such that the decomposition $\pmb{\rho} = \pmb{Z} \pmb{\Pi} \pmb{Z}^{\top} + \pmb{\Gamma}$  holds.

%--------------------------------------------------------
\subsection*{A.2. Proof of Theorem \ref{thm:global_minimum_variance}}
\label{sc3+}
We first notice that for any portfolio $P_J$ chosen by selecting one asset from each cluster in the coarsest partition $G^*$, the portfolio correlation matrix will be the same due to the block structure in the large correlation matrix among all assets.
Denote this portfolio correlation matrix by $\pmb{\rho}_P$.
Then the portfolio covariance matrix is
\[\pmb{\Sigma}_{P_J} = \pmb{V}_{P_J}\pmb{\rho}_P\pmb{V}_{P_J},\]
where
$\pmb{V}_{P_J}:=\text{diag}\left(\sqrt{\var(X_{J(1)})},\ldots,\sqrt{\var(X_{J(K)})}\right)$ is the diagonal matrix containing the standard deviation of all stocks in the portfolio.
Because we do not allow short selling in the minimum variance portfolio, the optimal weights $\pmb{w}^*$ falls into either of the following two cases.
\begin{enumerate}
    \item All weights in $\pmb{w}^*$ are positive, and $\pmb{w}^* = \pmb{\Sigma}_{P_J}^{-1}\pmb{1}/\left(\pmb{1}^\top\pmb{\Sigma}_{P_J}^{-1}\pmb{1}\right)$.
    \item Some weights in $\pmb{w}^*$ are zero, and the remaining weights are positive and satisfy $\pmb{w}^{*+} = (\pmb{\Sigma}_{P_J}^{+})^{-1}\pmb{1}/\left(\pmb{1}^\top(\pmb{\Sigma}_{P_J}^{+})^{-1}\pmb{1}\right)$, where $\pmb{\Sigma}_{P_J}^{+}$ represents the covariance matrix of stocks with positive weights in the optimal portfolio.
\end{enumerate}
We now show that, in both cases, the variance of the minimum variance portfolio is non-decreasing in all elements of $\pmb{V}_{P_J}$.

\quad In Case (1), where $\pmb{w}^* = \pmb{\Sigma}_{P_J}^{-1}\pmb{1}/\left(\pmb{1}^\top\pmb{\Sigma}_{P_J}^{-1}\pmb{1}\right)$, the variance of the portfolio is
\begin{equation}
    \var_{\min}(P_J) = \frac{1}{2\pmb{1}^\top\pmb{\Sigma}_{P_J}^{-1}\pmb{1}} = \frac{1}{2\pmb{1}^\top\pmb{V}_{P_J}^{-1}\pmb{\rho}_{P}^{-1}\pmb{V}_{P_J}^{-1}\pmb{1}}.
    \label{eq:var_of_min_var_portfolio}
\end{equation}
Let us disregard the $1/2$ scaling and focus on the denominator, which can be rewritten as\[
\sum_{i=1}^K \sum_{j=1}^K \rho_P^{-1}(i, j) a_i a_j,
\]where $\rho_P^{-1}(i, j)$ is the $j$-th element of the $i$-th row of $\pmb{\rho}_{P}^{-1}$, and $a_i:= 1/\sqrt{\var(X_{J(i)})}$.
The derivative w.r.t. $a_i$ is
\begin{equation}
    2\rho_P^{-1}(i, i)a_i + 2\sum_{j\neq i}\rho_P^{-1}(i, j)a_j.
    \label{eq:min_var_denominator_derivative}
\end{equation}
Because $\pmb{w}^* = \pmb{\Sigma}_{P_J}^{-1}\pmb{1}/\left(\pmb{1}^\top\pmb{\Sigma}_{P_J}^{-1}\pmb{1}\right) = \pmb{V}_{P_J}^{-1}\pmb{\rho}_{P}^{-1}\pmb{V}_{P_J}^{-1}\pmb{1}/\left(\pmb{1}^\top\pmb{\Sigma}_{P_J}^{-1}\pmb{1}\right)  \geq \pmb{0}$, we have for any $i\in[K]$,
\begin{equation*}
    \rho_P^{-1}(i, i)a_i^2 +\sum_{j\neq i}\rho_P^{-1}(i, j)a_i a_j \geq 0.
\end{equation*}
Dividing both sides by $a_i > 0$, we have\[
    \rho_P^{-1}(i, i)a_i + \sum_{j\neq i}\rho_P^{-1}(i, j)a_j \geq 0.
\]
This means that the derivative (\ref{eq:min_var_denominator_derivative}) is non-negative.
In other words, the denominator of (\ref{eq:var_of_min_var_portfolio}) is non-decreasing in $a_i$ for any $i\in[K]$, thus (\ref{eq:var_of_min_var_portfolio}) is non-decreasing in $\var(X_{J(i)})$ for any $i\in[K]$.

\quad Now let us consider Case (2).
If the optimal weight for a stock $X_i$ is positive, then, as we have shown in case (1), increasing $\var(X_i)$ will decrease $a_i$ and thus increase the variance of the optimal portfolio.
This also decreases its optimal weight $w^{*+}_i$, up to the point where $w^{*+}_i=0$.
If the optimal weight for a stock $X_i$ is zero, it means that, in the optimal minimum variance portfolio where short selling is allowed, the weight of $X_i$ is non-positive.
In other words,
\begin{equation}
    \rho_P^{-1}(i, i)a_i^2 +\sum_{j\neq i}\rho_P^{-1}(i, j)a_i a_j \leq 0.
    \label{eq:min_var_weight_numerator}
\end{equation}
Notice that the left-hand side of (\ref{eq:min_var_weight_numerator}) is quadratic in $a_i$, with one root at $a_i=0$ and another at some $a_i>0$.
Because the $\rho_P^{-1}(i, i)$, being the diagonal of the inverse correlation matrix, is larger than 1, increasing the variance of $X_i$ thus decreasing $a_i$ toward $0$ will only make (\ref{eq:min_var_weight_numerator}) stay negative. Hence the optimal weight of the no-short-selling problem will not change.
This shows that the variance of the minimum-variance portfolio is also non-decreasing in $\var(X_{J(i)})$ for any $i\in[K]$ in case (2).

%--------------------------------------------------------
\subsection*{A.3. Proof of Theorem \ref{thm:recovery}}
\label{sc33}
We start with the following lemma, which provides a sufficient condition under which the PARTITION algorithm recovers the partition $G^{\star}$.
\begin{lemma}
\label{lem:detrecovery}
Let $\tau:=\max_{i,j,l \in [d]} |(\widehat{\rho}_{il} - \widehat{\rho}_{jl}) - (\rho_{il} - \rho_{jl})|$, and $\Delta:= \min_{i \stackrel{G^{\star}}{\nsim} j} \cord(i,j)$.
If $\tau \le \varepsilon < \Delta - \tau$, then
the PARTITION algorithm with inputs $\widehat{\cord}$ and $\varepsilon$ outputs $\widehat{G} = G^{\star}$.
\end{lemma}
\begin{proof}
By the definition of $\tau$, we have $\widehat{\cord}(i,j)-\tau\leq \cord(i,j) \leq \widehat{\cord}(i,j) + \tau$.
If $i \stackrel{G^{\star}}{\sim} j$, we have $\cord(i,j) = 0$, and thus $\widehat{\cord}(i,j) \le \tau$.
Similarly, if $i \stackrel{G^{\star}}{\nsim} j$, we have $\cord(i,j) \ge \Delta$, and thus $\widehat{\cord}(i,j) \ge \Delta - \tau$.
Consequently, if $\tau \le \varepsilon < \Delta - \tau$, then
\begin{equation*}
i \stackrel{G^{\star}}{\sim} j  \quad \mbox{if and only if} \quad \widehat{\cord}(i,j) \le \varepsilon.
\end{equation*}

\quad Now we prove that the PARTITION algorithm recovers the partition $G^{\star}$.
We argue by induction, and assume that the algorithm is correct in the first $t-1$ steps:
\begin{equation*}
\widehat{G}_s = G^{\star}_{k(i_l)} \quad \mbox{for } l = 1, \ldots, t-1,
\end{equation*}
where $k(i_l)$ is the index of the group that contains $i_l$.
At step $t$, if $|S| = 1$, the algorithm terminates after this step and outputs $\widehat{G} = G^{\star}$.
If $|S| > 1$, there are two cases:
\begin{enumerate}[itemsep = 3 pt]
\item
If $\widehat{\cord}(i_t,j_t) > \varepsilon$, then $\widehat{\cord}(i_t, j) > \varepsilon$ and thus $i_t \stackrel{G^{\star}}{\nsim} j$ for any $j \in S$.
Since the algorithm is correct up to step $t-1$, $i_t \stackrel{G^{\star}}{\nsim} j$ for any $j \notin S$.
Thus, $i_t$ must be a singleton, and the algorithm outputs $\widehat{G}_t =  G^{\star}_{k(i_t)} = \{i_t\}$.
\item
If $\widehat{\cord}(i_t,j_t) \leq \varepsilon$, then $i_t \stackrel{G^{\star}}{\sim} j_t$.
The new cluster is $\widehat{G}_t = S \cap G_{k(i_t)}$.
Since the algorithm is correct in the first $t-1$ steps, $G_{k(i_t)} \subset S$.
Thus, $\widehat{G}_t = G_{k(i_t)}$.
\end{enumerate}
The PARTITION algorithm is correct in both cases at step $t$, which completes the induction.
\end{proof}

\quad The quantity $\tau$ is the sampling error of correlation differences.
Lemma \ref{lem:detrecovery} implies that if $\tau$ is small enough, the PARTITION algorithm recovers the partition $G^{\star}$ with a properly chosen $\varepsilon$.
To prove Theorem \ref{thm:recovery}, we also need the following lemma which gives an estimate of $\tau$.
\begin{lemma}
\label{lem:esttau}
Under Assumption \ref{assump:alphaexp}, there exist numerical constants $c_1, c_2 > 0$ such that
\begin{equation*}
\tau \le 2L^2 \left(c_1 \sqrt{\frac{\log d}{n}} + c_2 \frac{(\log d)^{\frac{2}{\alpha}}}{n} \right),
\end{equation*}
with probability at least $1-4/d$.
\end{lemma}

\quad A proof of Lemma \ref{lem:esttau} is based on the following result on the concentration of quadratic forms in i.i.d. random variables with $\alpha$-sub-exponential distribution.
Recall that $\pmb{X}^{*}$ is the $n \times d$ matrix whose row $r$ is $X^{*r} = (X^{*r}_1, \ldots, X^{*r}_d)$, the standardized returns in period $r \in [n]$.
\begin{lemma}
\label{lem:qfcon}
Under Assumption \ref{assump:alphaexp}, there exists $c > 0$ such that for any $t > 0$ and $u, v \in \mathbb{R}^d$,
we have
\begin{equation*}
\left| \frac{1}{n} u^{\top}\pmb{X}^{* \top}\pmb{X}^{*}v - u^{\top} \pmb{\rho} v  \right| \le (\ln{2})^{-2/\alpha} L^2 \sqrt{u^{\top} \pmb{\rho} u} \sqrt{v^{\top} \pmb{\rho} v}\left(c\sqrt{\frac{t}{n}} + c^{\frac{4}{\alpha}} \frac{t^{\frac{2}{\alpha}}}{n} \right),
\end{equation*}
with probability at least $1 - 4e^{-t}$.
\end{lemma}

\quad A proof of Lemma \ref{lem:qfcon} relies on the following concentration inequality of quadratic forms in i.i.d. random variables with $\alpha$-sub-exponential distribution.
\begin{lemma}
(\cite{GSS21, S20})
\label{lem:HWalpha}
For $\alpha \in (0,2]$, let $Y = (Y_1, \ldots, Y_n)$ be centered and independent random variables such that $||Y_r||_{\psi_{\alpha}} \le L$ for each $r \in [n]$.
Let $\pmb{A} = (a_{ij})_{i,j \in [n] \times [n]}$ be a symmetric matrix.
Then there exists a constant $C = C(\alpha)$ such that for any $t > 0$,
\begin{equation}
\label{eq:concenalpha}
\mathbb{P}\left(|Y^{\top} \pmb{A} Y - \mathbb{E}(Y^{\top} \pmb{A} Y) | > t \right) \le
2 \exp \left(-C \min \left(\frac{t^2}{L^4 ||\pmb{A}||_2^2}, \left(\frac{t}{L^2 ||\pmb{A}||_{op}} \right)^{\frac{\alpha}{2}} \right) \right),
\end{equation}
where $||\pmb{A}||_2:= \sqrt{\sum_{i,j = 1}^n a_{ij}^2}$ is the $2$-norm, and $||\pmb{A}||_{op}: = \sup\{|\pmb{A}x|: |x| = 1\}$ is the operator norm.
\end{lemma}

\quad See also \cite{RV13, A15, VW15, JL20} for related results on concentration inequalities of quadratic forms in i.i.d. random variables.
Now we proceed to prove Lemma \ref{lem:qfcon}.
\begin{proof}[Proof of Lemma \ref{lem:qfcon}]
First we express the inequality \eqref{eq:concenalpha} in a slightly different way.
By replacing $t$ with $s:=cL^2(||\pmb{A}||_2 \sqrt{t} + c^{\frac{4}{\alpha}-1} ||\pmb{A}||_{op} t^{\frac{2}{\alpha}})$ where $c>0$ is some constant and $t$ is any positive number, we have
\begin{align*}
\mathbb{P}(|Y^{\top} \pmb{A} Y - \mathbb{E}(Y^{\top} \pmb{A} Y) | \ge s) & \le 2 \exp \left(-C \min \left(\frac{s^2}{L^4 ||\pmb{A}||_2^2}, \left(\frac{s}{L^2 ||\pmb{A}||_{op}} \right)^{\frac{\alpha}{2}} \right) \right) \\
& \le 2 \exp \left(-C \min \left(\frac{(cL^2 ||\pmb{A}||_2 \sqrt{t})^2}{L^4 ||\pmb{A}||_2^2}, \left(\frac{cL^2 c^{\frac{4}{\alpha}-1} ||\pmb{A}||_{op} t^{\frac{2}{\alpha}}}{L^2 ||\pmb{A}||_{op}} \right)^{\frac{\alpha}{2}} \right) \right) \\
& = 2 \exp(-Cc^2 t).
\end{align*}
By taking $c = \sqrt{1/C}$, we get, for any $t>0$,
\begin{equation}
\label{eq:concenalpha2}
\mathbb{P}\left(\left|Y^{\top} \pmb{A} Y - \mathbb{E}\left(Y^{\top} \pmb{A} Y\right) \right| \ge cL^2\left(||\pmb{A}||_2 \sqrt{t} + c^{\frac{4}{\alpha}-1} ||\pmb{A}||_{op} t^{\frac{2}{\alpha}}\right)\right) \le 2 e^{-t}.
\end{equation}
By Assumption \ref{assump:alphaexp}, for any $\omega \in \mathbb{R}^d$ and $r \in [n]$, we have
$||(X^{*r})^{\top} \omega||_{\psi_{\alpha}} = ||(\pmb{\rho}^{-1/2}X^{*r})^{\top} (\pmb{\rho}^{1/2} \omega)||_{\psi_{\alpha}} \le ||\pmb{\rho}^{-1/2}X^{*r}||_{\psi_{\alpha}} ||\pmb{\rho}^{1/2} \omega||_{\psi_{\alpha}} \leq  L (\ln{2})^{-1/\alpha} \sqrt{\omega^{\top} \pmb{\rho} \omega}$.
By applying \eqref{eq:concenalpha2} to $Y = \pmb{X}^{*\top} \omega$ and $\pmb{A} = \pmb{I}_n$ with
$\omega = \lambda u  + \lambda^{-1}v$ and $\omega = \lambda u  - \lambda^{-1}v$ for some $\lambda > 0$, respectively, we get
\begin{align*}
&\Big||\lambda \pmb{X}^{*} u \pm \lambda^{-1}\pmb{X}^{*} v|^2 - \mathbb{E}\big[|\lambda \pmb{X}^{*} u \pm \lambda^{-1}\pmb{X}^{*} v|^2\big]\Big|
\\ \leq &(\ln{2})^{-2/\alpha}cL^2(\lambda u \pm \lambda^{-1} v)^{\top} \pmb{\rho}(\lambda u \pm \lambda^{-1} v)(\sqrt{nt}+c^{\frac{4}{\alpha}-1}t^{\frac{2}{\alpha}}),
\end{align*}
with probability at least $1 - 4 e^{-t}$.
Notice that $u^{\top}\pmb{X}^{* \top}\pmb{X}^{*}v = \frac{1}{4} (|\lambda \pmb{X}^{*} u + \lambda^{-1}\pmb{X}^{*} v|^2 - |\lambda \pmb{X}^{*} u - \lambda^{-1}\pmb{X}^{*} v|^2)$.
As a consequence,
\begin{align*}
\left| \frac{1}{n} u^{\top}\pmb{X}^{* \top}\pmb{X}^{*}v - u^{\top} \pmb{\rho} v  \right| & \le \frac{1}{4n}\Big( \Big||\lambda \pmb{X}^{*} u + \lambda^{-1}\pmb{X}^{*} v|^2 - \mathbb{E}\left[ |\lambda \pmb{X}^{*} u + \lambda^{-1}\pmb{X}^{*} v|^2\right] \Big| \\
& \qquad \qquad \qquad  + \Big||\lambda \pmb{X}^{*} u - \lambda^{-1}\pmb{X}^{*} v|^2 - \mathbb{E}\left[ |\lambda \pmb{X}^{*} u - \lambda^{-1}\pmb{X}^{*} v|^2\right] \Big| \Big) \\
& \le \frac{1}{4n}(\ln{2})^{-2/\alpha}cL^2(\sqrt{nt}+c^{\frac{4}{\alpha}-1}t^{\frac{2}{\alpha}})\Big[2\lambda^2u^{\top} \pmb{\rho} u +2\lambda^{-2}v^{\top}\pmb{\rho} v\Big].
\end{align*}
By taking $\lambda = (v^{\top} \pmb{\rho} v/u^{\top} \pmb{\rho} u)^{1/4}$, we get the desired result.
\end{proof}

\quad Now we prove Lemma \ref{lem:esttau}.
\begin{proof}[Proof of Lemma \ref{lem:esttau}]
For $i,j,l \in [d]$, let $u = e_i - e_j$ and $v = e_l$, where $e_i = (0, \ldots, 0,1,0, \ldots, 0)$ with the $i^{th}$ coordinate one and all others zeros.
It is easy to see that
\begin{equation*}
\widehat{\rho}_{il} - \widehat{\rho}_{jl} =  \frac{1}{n} u^{\top}\pmb{X}^{* \top}\pmb{X}^{*}v \quad \mbox{and} \quad
\rho_{il} - \rho_{jl} = u^{\top} \pmb{\rho} v.
\end{equation*}
By Lemma \ref{lem:qfcon}, we have for any $t > 0$,
\begin{align*}
|(\widehat{\rho}_{il} - \widehat{\rho}_{jl}) - (\rho_{il} - \rho_{jl})| & \le (\ln{2})^{-2/\alpha}L^2 \sqrt{2 - 2 \rho_{ij}} \left(c\sqrt{\frac{t}{n}} + c^{\frac{4}{\alpha}} \frac{t^{\frac{2}{\alpha}}}{n} \right) \\
& \le 2(\ln{2})^{-2/\alpha}L^2 \left(c\sqrt{\frac{t}{n}} + c^{\frac{4}{\alpha}} \frac{t^{\frac{2}{\alpha}}}{n} \right),
\end{align*}
with probability at least $1 - 4 e^{-t}$.
Notice that the above inequality holds for any $1 \le i < j \le d$ and $l \ne i, j$.
Taking $t = \log(d)$, we have
\begin{equation*}
    |(\widehat{\rho}_{il} - \widehat{\rho}_{jl}) - (\rho_{il} - \rho_{jl})| \le 2(\ln{2})^{-2/\alpha}L^2 \left(c\sqrt{\frac{\log d}{n}} + c^{\frac{4}{\alpha}} \frac{(\log d)^{\frac{2}{\alpha}}}{n} \right),
\end{equation*} for any $1 \le i < j \le d$ and $l \ne i, j$, with probability at least $1 - 4/d$.
Let $c_1 = (\ln{2})^{-2/\alpha}c$ and $c_2=(\ln{2})^{-2/\alpha}c^{4/\alpha}$, we have \[\tau=\max_{i,j,l \in [d]} |(\widehat{\rho}_{il} - \widehat{\rho}_{jl}) - (\rho_{il} - \rho_{jl})|\le 2L^2 \left(c_1\sqrt{\frac{\log d}{n}} + c_2 \frac{(\log d)^{\frac{2}{\alpha}}}{n} \right)\]with probability at least $1 - 4/d$.
\end{proof}

\quad Finally, Theorem \ref{thm:recovery} follows easily from Lemma \ref{lem:detrecovery} and Lemma \ref{lem:esttau}.

%--------------------------------------------------------
\subsection*{A.4. Proof of Theorem \ref{thm:complexity}}
\label{sc32}
The ACC algorithm is decomposed into two stages.

\quad The first is the preparation stage, where the sample correlation matrix and the $\widehat{\cord}$ matrix are computed, and the tail parameters are estimated.
The standardization step takes $\mathcal{O}(nd)$.
The complexity of computing $\widehat{\pmb \rho}$ is $\mathcal{O}(nd^2)$.
For any $i, j \in [d]$, the complexity of computing $\widehat{\cord}(i,j)$ is $\mathcal{O}(d)$.
So computing the entire $\widehat{\cord}$ matrix at most $d^2 \cdot \mathcal{O}(d) = \mathcal{O}(d^3)$ operations.
Calculating $\widehat{\pmb \rho}^{-1/2}$ can be done through eigenvalue decomposition, which has complexity $\mathcal{O}(d^3)$.
Sorting $Y_i$ takes $\mathcal{O}(n\log(n))$, and each linear regression takes $\mathcal{O}(n)$.
Therefore, the overall complexity of the preparation stage is $\mathcal{O}(d^2(n+d))$.

\quad The second stage is the grid search stage, where an appropriate $\varepsilon$ is chosen based on the results of the PARTITION procedure using different values for $\varepsilon$.
In the PARTITION procedure, the while loop has at most $d$ iterations.
In each iteration, finding $\argmin_{i,j \in S, i \ne j} \widehat{\cord}(i,j)$ would take $\mathcal{O}(d^3)$, but it can be simplified by keeping a sorted list of all values in $\widehat{\cord}$, since the same $\widehat{\cord}$ is passed to PARTITION every time.
With this sorted list, finding $\argmin_{i,j \in S, i \ne j} \widehat{\cord}(i,j)$ is at most $\mathcal{O}(d)$.
Finding the set $\left\{k \in S: \min\left(\pmb{D}(i_l, k), \pmb{D}(j_l, k)\right) \le \varepsilon \right\}$ is at most $\mathcal{O}(d)$, and all other steps are constant.
Overall, each PARTITION call takes $\mathcal{O}(d^2)$.
Sorting all entrees in $\widehat{\cord}$ in advance takes $\mathcal{O}(d^2\log(d^2))=\mathcal{O}(d^3)$.
Because the number of grids $n_g$ in the grid search is a constant and does not grow with $d$ or $n$, the grid search stage has complexity $\mathcal{O}(d^3)$.

\quad
Therefore, the overall complexity of Algorithm \ref{algo:ACC} is $\mathcal{O}(d^2(n+d))$.

\section*{Appendix B. Experimental Details}
\label{appendixB}

\quad In this section, we recall some algorithms used in our empirical study.

\subsection*{B.1. Single-linkage clustering}
\label{AppendixB1}
Single-linkage is an agglomerative hierarchical clustering method and follows the following general procedure  on a given distance matrix:
\begin{enumerate}
    \item Begin with $d$ clusters, each consisting of exactly one entity. Label the clusters with the numbers $1, \ldots, d$.
    \item Search the distance matrix for the closest pair of clusters. Let the chosen clusters be labeled $A$ and $B$, and their distance be $d_{A, B}$. This distance $d_{A, B}$ is usually based on distances between members of clusters $A$ and $B$.
    \item Merge clusters $A$ and $B$, thus reducing the total number of clusters by $1$. Relabel the merged cluster as $A$, and update the distance matrix to reflect the revised distance between cluster $A$ and all other existing clusters. Delete the row and column in the distance matrix pertaining to cluster $B$.
    \item Repeat Steps 2 and 3 until the desired number of clusters are obtained.
\end{enumerate}
\quad Different agglomerative hierarchical clustering algorithms have been proposed with different definitions of the distance measure $d_{A, B}$.
In the single-linkage, the distance between two clusters is defined as the distance between the two closest members of clusters $i$ and $j$: \[d_{A,B}:=\min_{a\in A,b\in B}D(a,b),\] where $D(a,b)$ is a given distance between entities $a$ and $b$.\footnote{Similar agglomerative hierarchical clustering algorithms include complete-linkage: $d_{A,B}:=\max_{a\in A,b\in B}D(a,b)$, and average-linkage: $d_{A,B}:=\frac{1}{|A||B|}\sum_{a\in A,b\in B}D(a,b)$.For a more detailed discussion we refer to \cite{ANDERBERG1973131}.}
In \cite{Man99}, Steps 2 and 3 in the single-linkage clustering is repeated until only one cluster remains.
By recording $\argmin_{a\in A,b\in B} D(a,b)$ in each merge, a maximum spanning tree is obtained, together with a hierarchical organization among all entities.

\subsection*{B.2. $k$-medoids clustering}
\label{AppendixB2}
The $k$-medoids clustering method \citep{Kaufman1990} aims to find clusters of similar entities by first identifying a set of $k$ representative entities. Then, $k$ clusters are constructed by assigning each entity to the nearest representative object. Such representative objects are called \textit{medoids} of the clusters, hence the name $k$-medoids. The $k$-medoids method can be formulated  as the following optimization problem:
\begin{align*}
    \min_{M}   \quad   &\sum_{i\in[d], i\notin M}\min_{m\in M}D(i, m)\\
    \text{subject to} \quad  &M\subset[d]\\
                        &|M|=k.
\end{align*}
Here, $M$ denotes the set of medoids, which must be a subset of all entities $[d]$ and must have cardinality $k$. $D(a,b)$ is a given distance between entities $a$ and $b$. For any given set of medoids $M$, we assign each non-medoid entity to its closest medoid $m$, and the optimization finds such a set of medoids that minimizes the shortest total distance between non-medoid entities and their corresponding medoids.

\quad The $k$-medoid is implemented as an iterative algorithm that gradually improves the quality of $M$. First, a set of initial medoids is chosen. Different initialization methods can be applied here. \cite{Kaufman1990} propose their own initialization, where $k$ medoids are selected in sequence such that the first medoid is the most centered entity, and each subsequent medoid decreases the objective function as much as possible. In practice, random initialization is often used for simplicity. In our implementation, the first medoid is randomly selected, and then each subsequent medoid is the entity with the largest distance from its closest existing medoid.\footnote{We use the python package \texttt{pyclustering} for this initialization and the subsequent $k$-medoid clustering.}

\quad After initialization, the algorithm improves the medoids by considering all possible swaps, i.e., replacing a medoid $h$ with a non-medoid entity $i$, and carrying out the swap that decreases the objective function as much as possible. The algorithm stops when no swap can decrease the objective function anymore.

\quad The distance measure between stocks $a$ and $b$ used in both the hierarchical clustering method and the $k$-medoids method is $D(a,b):= \sqrt{2(1-\rho_{ab})}$, the same as in \cite{Man99}, and the desired number of clusters $k$ is set to be $20$ for those two methods.

\subsection*{B.3. Risk parity}
\label{AppendixB3}
The concept of risk parity was pioneered by Bridgewater in its All Weather strategy launched in 1996.\footnote{``The All Weather Story'', Bridgewater, Accessed February 25, 2021. \url{https://www.bridgewater.com//_document?id=00000171-8623-d7de-affd-feaf4ee20000}}
Numerical experiments \citep{Maillard2010} show that risk parity often provides more balanced allocations, mitigating the problem of extreme portfolios in the mean-variance approach while providing better returns--risk tradeoff than equally-weighted portfolios.
In our experiments, we apply the version of risk parity that equalizes weighted marginal risk contribution, also adopted by \cite{Maillard2010}, of every asset in the portfolio.
To be more precise, define the volatility of a portfolio
\begin{equation}
\sigma(\pmb{w}) = \sqrt{\pmb{w}^T \pmb{\Sigma} \pmb{w}},
\end{equation}
where $\pmb{\Sigma}$ is the covariance matrix and $\pmb{w}$ is the vector of allocation weights of $d$ assets of the portfolio.
Hence, the risk contribution of asset $i$ is
\begin{equation}
\sigma_i(\pmb{w}) = w_i \frac{\partial \sigma(\pmb{w})}{\partial w_i} = \frac{w_i (\pmb{\Sigma} \pmb{w})_i}{\sigma(\pmb{w})}.
\end{equation}
Observe that $\sigma(\pmb{w}) = \sum_{i = 1}^d \sigma_i(\pmb{w})$.
We now construct a portfolio in such a way that the risk contribution of all assets equal, namely
\begin{equation}
\label{eq:rp}
\sigma_i(\pmb{w}) = \frac{\sigma(\pmb{w})}{d} \Longleftrightarrow w_i = \frac{\sigma(\pmb{w})^2}{d\cdot (\pmb{\Sigma} \pmb{w})_i}.
\end{equation}
It is easy to see that the problem \eqref{eq:rp} is equivalent to the non-linear optimization problem
\begin{equation}
\begin{aligned}
 \min_{\pmb{w}} \quad &\sum_{i = 1}^d \left(w_i - \frac{\sigma(\pmb{w})^2}{d\cdot (\pmb{\Sigma} \pmb{w})_i} \right)^2 \\
\text{subject to}\quad &\pmb{w}^\top \pmb{1} = 1, \, \pmb{w}\geq 0.
\end{aligned}
\end{equation}

\subsection*{B.4. Markowitz's mean-variance strategy and minimum variance strategy}
\label{AppendixB4}
Markowitz's original mean-variance strategy without short-selling:
\begin{align}
    \min_{\pmb{w}} \quad &\pmb{w}^\top \pmb{\Sigma}\pmb{w}\\
    \text{subject to}\quad &\pmb{w}^\top \pmb{\mu}\geq\alpha\nonumber\\
    &\pmb{w}^\top \pmb{1} = 1, \, \pmb{w}\geq 0, \nonumber
\end{align}
where $\pmb{\mu}\in\mathbb{R}^d$ contains the mean returns of the stocks, which is estimated using the average of the daily returns in the backward-looking window, and $\alpha$ is the target return, which we set to 10\%.

\quad The minimum variance strategy is similar to Markowitz's mean-variance optimization but without the expected return constraint:
\begin{align}
    \min_{\pmb{w}} \quad &\pmb{w}^\top \pmb{\Sigma}\pmb{w}\\
    \text{subject to} \quad &\pmb{w}^\top \pmb{1} = 1, \, \pmb{w}\geq 0. \nonumber
\end{align}

\bibliographystyle{abbrvnat}
\bibliography{unique}
\end{document}